\documentclass[11pt]{article}
\usepackage{amsfonts}
\usepackage{amssymb}
\usepackage{amscd}
\newcommand{\C}{\mathbb C}
\newcommand{\G}{\mathbb G}
\newcommand{\bp}{{\mathbb P}}
\newcommand{\R}{\mathbb R}

\newcommand{\la}{\langle}
\newcommand{\ra}{\rangle}
\newcommand{\ca}{{\mathcal A}}
\newcommand{\cb}{{\mathcal B}}

\newcommand{\ce}{{\mathcal E}}
\newcommand{\cf}{{\mathcal F}}
\newcommand{\cg}{{\mathcal G}}
\newcommand{\ch}{{\mathcal H}}
\newcommand{\cj}{{\mathcal J}}
\newcommand{\cl}{{\mathcal L}}
\newcommand{\rp}{{\mathcal P}}
\newcommand{\cm}{{\mathcal M}}

\newcommand{\co}{{\mathcal O}}
\newcommand{\cs}{{\mathcal S}}
\newcommand{\ct}{{\mathcal T}}
\newcommand{\cv}{{\mathcal V}}
\newcommand{\U}{{\mathcal {U}}}
\newcommand{\cw}{{\mathcal W}}
\newcommand{\cz}{{\mathcal Z}}
\newcommand{\cp}{\C{\bp}^1_x}

\newcommand{\ad}{{\mathop{\mbox{Ad}}\nolimits}}

\newcommand{\im}{{\mathop{\mbox{Im}}\nolimits\,}}
\renewcommand{\ker}{{\mathop{\mbox{Ker}}\nolimits\,}}

\newcommand{\fm}{{\mathfrak M}}

\newcommand{\fh}{{\mathfrak H}}
\newcommand{\fj}{{\mathfrak j}}
\newcommand{\fg}{{\mathfrak g}}
\newcommand{\fo}{{\mathfrak O}}
\newcommand{\fs}{{\mathfrak S}}
\newcommand{\fu}{{\mathfrak U}}
\newcommand{\fx}{{\mathfrak X}}
\def\a{\alpha}
\def\b{\beta}
\def\g{\gamma}
\def\d{\delta}

\def\l{\lambda}
\def\o{\omega}
\def\p{\partial}
\def\ve{\varepsilon}
\def\vp{\varphi}
\def\r{\rho}

\def\t{\triangle}

\def\bl{\backslash}
\def\hra{\hookrightarrow}
\def\lra{\longrightarrow}

\begin{document}
\begin{flushright}
 hep-th/9806239
\end{flushright}
\vskip 3cm

\begin{center}
{\LARGE \bf Holomorphic Chern-Simons-Witten Theory:\\
\vskip 0.5cm
from 2D to 4D Conformal Field Theories}
\vskip 1.5cm
{\Large A.D.Popov}\footnote {E-mail: popov@thsun1.jinr.dubna.su}\\
\vskip 1cm
{\em Bogoliubov Laboratory of Theoretical Physics\\
JINR, 141980 Dubna, Moscow Region, Russia}\\

\vskip 1.1cm
\end{center}

\begin{quote}
{\bf Abstract.}
It is well known that rational 2D conformal field theories
are connected with Chern-Simons theories defined on 3D {\it real}
manifolds. We consider holomorphic analogues of Chern-Simons
theories defined on 3D {\it complex} manifolds (six real dimensions)
and describe 4D conformal field theories connected with them.
All these models are integrable. We describe analogues of the Virasoro
and affine Lie algebras, the local action of which on  fields of
holomorphic analogues of Chern-Simons theories becomes nonlocal
after pushing down to the action on fields of integrable 4D conformal
field theories. Quantization of integrable 4D conformal field theories
and relations to string theories are briefly discussed.
\end{quote}

\newpage
\tableofcontents
\newpage

\section{Introduction}
\label{1}

Rational 2D conformal field theories (CFT) form an important subclass
of CFT's in two dimensions. They are completely solvable and may be
obtained from 3D Chern-Simons theories by an appropriate choice of a
gauge group and coupling constants~\cite{Wit1}-\cite{EMSS}. The aim of
the present paper is to describe 4D CFT's, which are as solvable as
rational 2D CFT's, and successful nonperturbative quantization of
which may be developed. These integrable 4D CFT's are connected
with holomorphic analogues of Chern-Simons
theories~\cite{Wit2}-\cite{Po} in the same way as rational 2D CFT's
are related to ordinary Chern-Simons theories. This correspondence
between 4D and 6D theories is different from the AdS/CFT
correspondence being extensively discussed in the literature. We
also consider free 4D CFT's describing fields of arbitrary spin
$s=0,1/2,1,...$ (and negative helicity) on 4D manifolds with
the self-dual Weyl tensor.

\smallskip

In many respects the progress in 2D CFT's was related to the
existence of infinite-dimensional symmetry algebras among which the
Virasoro and affine Lie algebras play the most important role. In
contrast with significant progress in understanding 2D CFT's the
knowledge of 4D CFT's is much less explicit, and not so many exact
results are obtained (see e.g.~\cite{Mack}-\cite{APSW} and
references therein). Usually, one connects this with the fact that,
unlike the 2D case, the conformal group in four dimensions (which
serves for the classification of primary fields in the limit
of flat 4D space~\cite{MaS}) is finite-dimensional, and constraints
arising from conformal invariance are not sufficient for a detailed
description of 4D CFT's. This is true for the general case, but for
a special subclass of 4D CFT's - integrable 4D CFT's - this is a
wrong impression based on the consideration of only local (manifest)
symmetries.

\smallskip

Integrable 4D CFT's are related to self-dual theories, which was
discussed in the paper~\cite{Po}.
By now it is well known that self-duality in four dimensions is
connected with holomorphy in six dimensions~\cite{Pen}-\cite{H1}.
Correspondingly, 4D self-duality/6D holomorphy and symmetry algebras
of 4D self-dual theories are the replacement of 2D self-duality/2D
holomorphy and 2D conformal invariance. The
discussion of this topic in the present paper is based on the
results of~\cite{Po} and much of the notation and terminology is
taken directly from that work. In Appendices we recall
definitions of those notions which we shall use.

\section{Chern-Simons theories}
\label{2}

\subsection{Definitions and notation}
\label{2.1}

Let $X$ be an oriented real 3D smooth manifold, $G$ a semisimple
compact Lie group, $\fg$ its Lie algebra, $P$ a principal $G$-bundle
over $X$, $A$ a connection 1-form on $P$ and $F(A)=dA + A\wedge A$
its curvature. Locally $A$ is a $\fg$-valued 1-form on $X$, and $F(A)$
is the $\fg$-valued 2-form on $X$.

\smallskip

Suppose a representation of $G$ in the complex vector space $\C^n$
is given. We associate  the complex vector bundle $\ce = P\times_G\C^n$
with the principal bundle $P$, and we shall use both the principal
bundle $P$ and the vector bundle $\ce$ in the description of
Chern-Simons theories.

\smallskip

Chern-Simons (CS) theories (see e.g.~\cite{Wit1}-\cite{EMSS},\cite
{DJT}-\cite{Fr} and references therein) describe locally constant
$G$-bundles, that is, bundles with locally constant transition
matrices. Put another way, a locally constant $G$-bundle is the
bundle with a flat connection $A$ and the field equations of the
CS theory are
$$
{}F(A) = dA + A\wedge A = 0.
\eqno(2.1)
$$
Locally eqs.(2.1) are solved trivially, and on any sufficiently
small open set $U\subset X$ we have $A = \phi^{-1}d\phi$, where
$\phi (x)\in G$. So, locally $A$ is a pure gauge.

\subsection{Moduli space of flat connections}
\label{2.2}

We denote by $\mathbb A$ the space of (irreducible) connections on
$P$ and by $\cg$ the infinite-dimensional group of gauge
transformations $A\mapsto A^g = g^{-1}Ag + g^{-1}dg$ (automorphisms
of $P$ which induce the identity transformation of $X$). Then the
moduli space $\fm_X$ of flat connections on $P$ is the space of
solutions to eqs.(2.1) modulo gauge transformations,
$$
\fm_X = \left\{ A\in {\mathbb A}: F(A) = 0 \right\}/\cg .
\eqno(2.2)
$$
In other words, the moduli space $\fm_X$ of flat connections on
$P$ (and on the vector bundle $\ce$) is the space of gauge
nonequivalent solutions to eqs.(2.1).

\smallskip

There is another description of the moduli space $\fm_X$ as the
space of homomorphisms from the fundamental group $\pi_1(X)$ of the
manifold $X$ to the group $G$ modulo conjugation by $G$,
$$
\fm_X = {\mbox{Hom}}(\pi_1(X),G)/G .
\eqno(2.3)
$$
Just this description of the moduli space of flat connections on
$\ce$ was used most often in papers on CS theories. For technical
complications in the case when $X$ is not connected, see e.g.~\cite{Fr}.

\smallskip

Below we want to consider the sheaf description of locally constant
bundles $\ce\to X$ from which  both (2.2) and (2.3) descriptions of
the moduli space $\fm_X$ of flat connections on $\ce$ follow. This
description is known in mathematics (see e.g.~\cite{Oni} and
references therein), and it can be compared with the description of
holomorphic bundles over complex manifolds. This will be useful for
introducing a holomorphic analogue of the CS theory. For simplicity
one usually considers locally constant bundles $\ce$ which are
equivalent to the trivial one as smooth vector bundles (if one
takes connected and simply connected structure groups $G$). The
generalization to topologically nontrivial bundles is straightforward
(see e.g.~\cite{MS},~\cite{DW}-\cite{Oni}). In \S\,2.3, we shall
introduce all necessary sheaves, and in \S\,2.4 we shall describe
moduli of flat connections in terms of cohomology sets. For
definitions of sheaves, cohomology sets etc, see Appendices.

\subsection{Sheaves $\mathbb G$, $\fs$ and $\ca$}
\label{2.3}

Let us consider the sheaf $\fs$ (of germs) of {\it smooth} maps
{}from $X$ into the group $G$ and its subsheaf
$\mathbb G:=X\times G$, continuous sections of which are locally
 constant maps from $X$ into $G$. Sections of the sheaf $\fs$ over
 an open set $U\subset X$ are smooth $G$-valued functions
 $\phi\in \fs (U)$ on $U$. Consider also the sheaf $\ca^q$ of smooth
q-forms on $X$ with values in the Lie algebra $\fg$ (q = 0,1,...).
The space of sections of the sheaf $\ca^q$ over an open set
$U\subset X$ is the space $\ca^q(U)$ of smooth $\fg$-valued
q-forms on $U$.

\smallskip

The sheaf $\fs$ acts on the sheaves $\ca^q$, q = 0,1,..., with the
help of the adjoint representation $\ad$. In particular, for any open
set $U\subset X$ we have
$$
A\mapsto \ad_\phi A=\phi^{-1}A\phi + \phi^{-1}d\phi ,
\eqno(2.4a)
$$
$$
{}F\mapsto \ad_\phi F=\phi^{-1}F\phi  ,
\eqno(2.4b)
$$
where $\phi \in \fs (U)$, $A\in \ca^1(U)$, $F\in \ca^2(U)$.

\smallskip

Denote by $i : \mathbb G\to\fs$ an embedding of $\mathbb G$ into
$\fs$. We define a map $\d^0 : \fs \to \ca^1$ given for any open
set $U$ of the space $X$ by the formula
$$
\d^0\phi = -(d\phi )\phi^{-1},
\eqno(2.5)
$$
where $\phi \in \fs (U)$, $\d^0 \phi \in \ca^1(U)$. Let us also
introduce an operator $\d^1 : \ca^1 \to \ca^2$, defined for an open
 set $U\subset X$ by the formula
$$
\d^1A = dA + A\wedge A ,
\eqno(2.6)
$$
where $A\in \ca^1(U)$, $\d^1A\in \ca^2(U)$. In other words, the maps
of sheaves $\d^0 : \fs\to \ca^1$ and $\d^1 : \ca^1\to\ca^2$ are
defined by means of localization. Finally, we denote by $\ca$ the
subsheaf in $\ca^1$, consisting of locally defined $\fg$-valued
1-forms $A$ such that $\d^1A = 0$, i.e. sections $A\in \ca (U)$
of the sheaf $\ca =\ker \d^1$ over $U\subset X$ satisfy eqs.(2.1).

\subsection{Locally constant bundles and cohomology sets}
\label{2.4}

It is not difficult to verify that the sequence of sheaves
$$
{\bf 1}\lra\G\stackrel{i}{\lra}\fs\stackrel{\d^0}{\lra}
\ca\stackrel{\d^1}{\lra}0
\eqno(2.7)
$$
is exact. From (2.7) we obtain the exact sequence of cohomology
sets~\cite{Oni}
$$
e\lra H^0(X,\G)\stackrel{i_*}{\lra}H^0(X,\fs) \stackrel
{\d^0_*}{\lra}H^0(X,\ca)\stackrel{\d^1_*}{\lra}
H^1(X,\G)\stackrel{\r}{\lra} H^1(X,\fs) ,
\eqno(2.8)
$$
where $e$ is a marked element (identity) of the considered sets, and
the  map $\r$ coincides with the canonical embedding
induced by the embedding of sheaves $i: \G\to\fs$.

\smallskip

By definition, (all) locally constant $G$-bundles $\ce$ over $X$
are parametrized by the set $H^1(X,\G)$. The kernel $\ker\r =
\r^{-1}(e)$ of the map $\r$ coincides with a subset of those
 elements from $H^1(X,\G)$ which are mapped into the class
$e\in H^1(X,\fs)$ of smoothly trivial bundles. By virtue of the
exactness of the sequence (2.8), the space $\ker \r$ is bijective to
the quotient space $H^0(X,\ca)/H^0(X,\fs)$. But $H^0(X,\ca)$ is
the space of global solutions to eqs.(2.1), $H^0(X,\fs)$ coincides
with the group of gauge transformations $\cg$, and therefore
$H^0(X,\ca)/H^0(X,\fs)$ coincides with the moduli space $\fm_X$ of
{}flat connections on $\ce$. So we have
$$
\fm_X = H^0(X,\ca)/H^0(X,\fs) \simeq \ker \r \subset H^1(X,\G) ,
\eqno(2.9)
$$
i.e. there is a one-to-one correspondence between the moduli space
 $\fm_X$ of flat connections on $\ce$ and the moduli space
$\ker\r =\r^{-1}(e)\subset H^1(X,\G)$ of those locally constant
bundles $\ce\to X$ which are trivial as smooth bundles.

\smallskip

{\bf Remark.}  One may also consider deformations of a locally
constant bundle $\ce\to X$ which is not topologically trivial.
If this bundle is associated with a principal bundle $P$ and
parametrized as a smooth bundle by a modulus $p\in H^1(X,\fs )$,
then instead of the sheaf $\ca^1$ one should take the sheaf
of 1-forms with values in the bundle $\ad P=P\times_G\fg$ and instead
of the sheaf $\fs$ one should take the sheaf
$\fs '$ of sections of the bundle Int$\,P$ (the bundle of groups
$P\times_GG$ where $G$ acts on itself by conjugation). Then
we shall have exact sequences analogous to (2.7), (2.8) but
with these new sheaves, and the point $e\in H^1(X,\fs ')$
will correspond to the point $p\in H^1(X,\fs )$.

\smallskip

If the structure group $G$ of the bundle $P$ is connected
and simply connected, then $H^1(X,\fs )=e$ and therefore
$\ker\r = H^1(X,{\mathbb G})$. That is, for topologically trivial
$G$-bundles we have
$$
H^0(X,\ca)/H^0(X,\fs) \simeq H^1(X,\G).
\eqno(2.10)
$$
Bijection (2.10) is a non-Abelian variant of the isomorphism between
\v{C}ech and de~Rham cohomologies.

\subsection{Rational 2D conformal field theories}
\label{2.5}

It is well known that 3D Chern-Simons theories on $X$
are connected with 2D conformal field theories if one supposes
that a 3-manifold $X$ has the form $\Sigma\times \R$
(or $\Sigma\times S^1$), where $\Sigma$ is a 2-manifold with or
without a boundary~\cite{Wit1}-\cite{EMSS}. Notice that the moduli space
$\fm_{\Sigma\times\R}$ of flat connections on a $G$-bundle over
$\Sigma\times\R$ is bijective to the moduli space $\fm_{\Sigma}$
of flat connections on the induced $G$-bundle over $\Sigma$ (for proof
see e.g.~\cite{Fr}). Moreover, if $\Sigma$  has a boundary, the
quantum Hilbert space $H_{\Sigma}$  is an infinite-dimensional
representation space of the chiral algebra of a CFT on $\Sigma$.
In particular, if $\Sigma = D_0$ is a disk (e.g. $|z|\le 1$) on the
complex plane $\C$, then the CS theory on $D_0\times\R$ reproduces the
chiral (holomorphic) version of the WZNW model~\cite{Wit1}-\cite{EMSS}.
The phase space of this model is the space of based loops $LG/G$.

\smallskip

The specification of a 3D CS theory is performed by a choice of
gauge group $G$, coupling constants and a manifold $X$. The
transition to 2D CFT is carried out by a choice of a 2-manifold
$\Sigma$ (disk, annulus, torus,...) in the splitting
 $X=\Sigma \times \R$ (or $X=\Sigma \times S^1$) and by a choice
of sources on $\Sigma$~\cite{Wit1}-\cite{EMSS},\cite{DJT}-\cite{Fr}.
{}For instance, we recover the above-mentioned chiral WZNW theory
without sources if we choose a {\it  connected} and
{\it simply connected} gauge group $G$ and $X = D_0\times \R$. If
we take {\it connected} but {\it non-simply connected} groups $G$,
we obtain a special subclass of rational 2D CFT's~\cite{MS,DW}.
One can also consider {\it disconnected} groups of type $N\ltimes G$,
where $G$ is a connected group with a discrete automorphism group $N$,
which leads to so-called {\it orbifold models}~\cite{MS, DW}.

\smallskip

At last, an important subclass of 2D CFT's is formed by $G/H$
{\it coset models} resulting from CS theories with a gauge
group $G\times H$ when coupling constants of gauge fields of the groups
$G$ and $H$ have opposite signs~\cite{MS}. For a more detailed discussion,
see~\cite{Wit1}-\cite{EMSS},\cite{DJT}-\cite{Fr}.

\section{Holomorphic Chern-Simons-Witten theories}
\label{3}

\smallskip

\subsection{Definitions and notation}
\label{3.1}

Let us consider a smooth six-dimensional manifold $\cz$ with an
integrable almost complex structure $\cj$. Then $\cz$ is a complex
3-manifold, and one can introduce a cover $\fu =\{\U_\a\}$ of $\cz$ and
complex coordinates $z_\a :\U_\a\to \C^3$. Let ${\bf G}=G^{\C}$ be a
complex semisimple Lie group, ${\bf g}=\fg^{\C}$ its Lie algebra, $P'$
a principal  ${\bf G}$-bundle over $\cz$ and $E'=P'\times_{\bf G}\C^n$
a smooth complex
vector bundle of rank $n$ associated with $P'$.

\smallskip

Let $\hat B$ be the
(0,1)-component of a connection 1-form on the bundle $E'$.  Suppose
that $\hat B$ satisfies the equations
$$
\bar\p\hat B + \hat B\wedge\hat B =0,
\eqno(3.1)
$$
where $\bar\p$ is the (0,1)-part of the exterior derivative
$d=\p+\bar\p$.  Equations (3.1) mean that the (0,2)-part of
the curvature $F$ of the bundle $E'$ is equal to
zero: $F^{0,2}:=\bar \p^2_{\hat B}=(\bar\p +\hat B)^2=0$ and
therefore the bundle $E'$ is holomorphic. We shall call eqs.(3.1)
 the field equations of
holomorphic Chern-Simons-Witten (CSW) theory.
This theory describes holomorphic bundles over a complex 3-manifold
$\cz$. We shall also use the abbreviation `hCSW theory' for it.

\smallskip

Equations (3.1) were suggested by Witten~\cite{Wit2} for a special
case of bundles over Calabi-Yau (CY) 3-folds $\cz$ as equations of a
holomorphic analogue of the ordinary Chern-Simons theory. Witten obtained
eqs.(3.1) from open $N=2$ topological strings with a central charge
$\hat c=3$ (6D target space), and the CY restriction $c_1(\cz)=0$  arised
{}from $N=2$ superconformal invariance of a sigma model used in constructing
the topological string theory. The connection of eqs.(3.1) with
topological strings was also considered in~\cite{BCOV}.  For the
hCSW theory on a CY 3-fold $\cz$, eqs. (3.1) follow from the action
~\cite{Wit2, BCOV}
$$
S_0 = \frac{1}{2}\int_{\cz}\hat\Omega\wedge\mbox{Tr}(\hat B\wedge
\bar\p\hat B + \frac{2}{3}\hat B\wedge\hat B\wedge \hat B),
\eqno(3.2)
$$
where $\hat\Omega$ is a nowhere vanishing holomorphic (3,0)-form on $\cz$.
The theory (3.1), (3.2) describes nonequivalent holomorphic
structures  on the bundle $E'\to\cz$.

\smallskip

Equations (3.1) on CY 3-folds were considered by Donaldson and
Thomas~\cite{DT} in the frames of the programme on extending the results
of Casson, Floer and Donaldson to manifolds of dimension D$>$4.
Donaldson and Thomas~\cite{DT} pointed out that one may
consider a more general situation with eqs.(3.1) on complex manifolds
$\cz$ which are not Calabi-Yau. We shall consider this more general
case of bundles over arbitrary complex 3-manifolds.

\smallskip

\subsection{Sheaves $\hat\ch$, $\hat\cs$ and $\hat\cb$}
\label{3.2}

To describe the moduli space of the holomorphic CSW theory
as it has been done in \S\,2  for the ordinary CS theory,
we should introduce some new sheaves and maps of sheaves.
{}For simplicity, we shall consider the smooth deformation of the
trivial holomorphic bundle $\cz\times\C^n$ over $\cz$.
Generalization to the case of topologically nontrivial bundles
is straightforward and not difficult.

\smallskip

Consider the sheaf $\hat\cs$ of {\it smooth} maps from $\cz$
into the group $\bf G$. The sheaf $\hat\ch$ of {\it holomorphic}
maps $\cz\to \bf G$ is a subsheaf of the sheaf $\hat\cs$, and there
exists a canonical embedding $i: \hat\ch\to\hat\cs$.
Consider also the sheaf $\hat\cb^{0,q}$ of {\it
smooth}  (0,q)-forms on $\cz$ with values in the Lie algebra
$\bf g$.
The sheaf $\hat\cs$ acts on the sheaves $\hat\cb^{0,q}$ ($q=1,2,...$)
with the help of the adjoint representation.  In particular, for any
open set $\U\subset\cz$ we have
$$
\hat B\mapsto \mbox{Ad}_{\hat\psi} \hat B =
\hat\psi^{-1}\hat B \hat\psi + \hat\psi^{-1}\bar\p\hat\psi ,
\eqno(3.3a)
$$
$$
\hat F\mapsto \mbox{Ad}_{\hat\psi} \hat F =
\hat\psi^{-1}\hat F \hat\psi ,
\eqno(3.3b)
$$
where $\hat\psi\in\hat\cs(\U)$, $\hat B\in \hat\cb^{0,1}(\U)$,
$\hat F\in \hat\cb^{0,2}(\U)$.

\smallskip

Let us define a map $\bar\delta^0:
\hat\cs\to\hat\cb^{0,1}$ given for any open set $\U$ of the space
$\cz$ by the formula
$$
\bar\d^0 \hat\psi = -(\bar\p \hat\psi )\hat\psi^{-1},
\eqno(3.4)
$$
where $\hat\psi\in\hat\cs(\U)$, $\bar\d^0 \hat\psi \in \hat\cb^{0,1}(\U)$,
$d=\p +\bar\p$. Let us also introduce an operator $\bar\d^1:  \hat\cb^{0,1}
\to\hat\cb^{0,2}$, defined for any open set $\U\subset\cz$ by the formula
$$
\bar\d^1\hat B =\bar\p\hat B + \hat B\wedge\hat B,
\eqno(3.5)
$$
where $\hat B\in \hat\cb^{0,1}(\U)$, $\bar\d^1\hat B\in
\hat\cb^{0,2}(\U)$. In other words, the maps of sheaves $\bar\d^0:
\hat\cs\to \hat\cb^{0,1}$ and $\bar\d^1: \hat\cb^{0,1} \to\hat\cb^{0,2}$
are defined by means of localizations.

\smallskip

Denote by $\hat\cb$ the subsheaf in  $\hat\cb^{0,1}$ consisting (of
germs) of (0,1)-forms $\hat B$ with values in $\bf g$ such that
$\bar\d^1\hat B=0$, i.e. sections $\hat B$ over any open set $\U$
of the sheaf $\hat\cb=\ker\bar\d^1$ satisfy the equations
$$
F^{0,2}(\hat B)= \bar\p\hat B + \hat B\wedge\hat B=0,
\eqno(3.6)
$$
where $\hat B\in \hat\cb^{0,1}(\U)$. So the sheaf $\hat\cb$ can be
identified with the sheaf of flat (0,1)-connections
$\bar\p_{\hat B}=\bar\p+\hat B$ on the complex vector bundle $E'$
over $\cz$ (and on the principal bundle $P'$).

\smallskip

\subsection{Moduli space of flat (0,1)-connections }
\label{3.3}

Let us consider the sheaves $\hat\cs$, $\hat\cb^{0,1}$ and
$\hat\cb^{0,2}$.
The triple $\{\hat\cs, \hat\cb^{0,1}, \hat\cb^{0,2}\}$ with the maps
$\bar\d^0$ and $\bar\d^1$ is a resolution of the sheaf $\hat\ch$, i.e.
the sequence of sheaves
$$
{\bf 1}\lra\hat\ch\stackrel{i}{\lra}\hat\cs\stackrel{\bar\d^0}{\lra}
\hat\cb^{0,1}
\stackrel{\bar\d^1}{\lra}\hat\cb^{0,2},
\eqno(3.7)
$$
where $i$ is an embedding, is exact~\cite{Oni}.

\smallskip

By virtue of the exactness of the sequence  (3.7), we have
$$
\im\bar\d^0= \ker\bar\d^1=\hat\cb .
\eqno(3.8)
$$
Since $\bar\d^0$ is the projection connected with the action (3.3a)
of the sheaf $\hat\cs$ on  $\hat\cb^{0,1}$, the sheaf $\hat\cs$  acts
transitively with the help of Ad on $\hat\cb$ and
$\hat\cb\simeq\hat\cs/\hat\ch$.  Thus, we obtain the exact sequence of
sheaves
$$
{\bf 1}\lra\hat\ch\stackrel{i}{\lra}\hat\cs\stackrel{\bar\d^0}{\lra}
\hat\cb\stackrel{\bar\d^1}{\lra}0.
\eqno(3.9)
$$
{}From (3.9) we obtain the exact sequence of cohomology sets
$$
e\lra H^0(\cz,\hat\ch)\stackrel{i_*}{\lra}H^0(\cz,\hat\cs) \stackrel
{\bar\d^0_*}{\lra}H^0(\cz,\hat\cb)\stackrel{\bar\d^1_*}{\lra}
H^1(\cz,\hat\ch)\stackrel{\hat\vp}{\lra} H^1(\cz,\hat\cs) ,
\eqno(3.10)
$$
where $e$ is a marked element (identity) of the considered sets, and
the map $\hat\vp$ coincides with the canonical embedding,
induced by the embedding of sheaves $i: \hat\ch\to\hat\cs$.

\smallskip

There is a one-to-one correspondence between the set $H^1(\cz,\hat\ch)$
and the set of equivalence classes (moduli space) of holomorphic
$\bf G$-bundles over $\cz$. The kernel
$\ker\hat\vp=\hat\vp^{-1}(e)$ of the map $\hat\vp$ coincides with a
subset of those elements from $H^1(\cz,\hat\ch)$, which are mapped into
the class $e\in H^1(\cz,\hat\cs)$ of topologically (and smoothly)
trivial bundles.  This means that representatives of the subset
$\ker\hat\vp$ are those transition matrices $\cf\in Z^1(\fu,\hat\ch)$ for
which there exists a splitting
$$
\cf_{\a\b}=\psi_{\a}^{-1}\psi_{\b}
\eqno(3.11)
$$
with {\it smooth} matrix-valued functions $\{\psi_\a\} \in
C^0(\fu, \hat S)$. Here $C^0(\fu, \hat\cs)$ is a group of 0-cochains
with the coefficients in $\hat\cs$, and
$Z^1(\fu,\hat\ch)$ is a set of 1-cocycles with the coefficients
in $\hat\ch$ (for definitions, see Appendices).

\smallskip

The moduli space $\cm_{\cz}$ of flat (0,1)-connections on $P'$
(and on $E'$) is the space of gauge nonequivalent global solutions
to eqs.(3.1). By definition we have
$\cm_{\cz}=H^0(\cz,\hat\cb)/H^0(\cz,\hat\cs)$, where
$H^0(\cz,\hat\cb)$ is the space of global solutions to eqs.(3.1)
and $H^0(\cz,\hat\cs)$ is the group of gauge transformations of
the hCSW theory. From the exactness of the sequence (3.10) we obtain
$$
\cm_{\cz}=H^0(\cz,\hat\cb)/H^0(\cz,\hat\cs)\simeq
\ker\hat \vp \subset H^1(\cz,\hat\ch),
\eqno(3.12)
$$
i.e. the moduli space $\cm_{\cz}$ of global
solutions of eqs.(3.6) on $\cz$ is bijective to the moduli space of
holomorphic bundles over $\cz$ which are trivial as smooth bundles.
Transition matrices of such bundles have the form (3.11).

\smallskip

{\bf Remarks}

\smallskip

{\bf 1.} Using the sheaves $\hat\cs$ and $\hat\cb$, considered in
\S\,3.2, one can introduce a {\it Dolbeault 1-cohomology set}
$H^{0,1}_{\bar\p_{\hat B}}(\cz)$
as a set of orbits of the group $H^0(\cz,\hat\cs)$ in the set
$H^0(\cz,\hat\cb)$, i.e.
$$
H^{0,1}_{\bar\p_{\hat B}}(\cz):= H^0(\cz,\hat\cb)/H^0(\cz,\hat\cs).
\eqno(3.13)
$$
 It follows from (3.12)
 that $H^{0,1}_{\bar\p_{\hat B}}(\cz)\simeq \ker\hat\vp$, and this
is a non-Abelian variant of the isomorphism between \v{C}ech and
Dolbeault cohomologies. Notice that deformations of a topologically
nontrivial holomorphic vector bundle $E'\to \cz$ can be described
by introducing  sheaves of sections of the bundles $\ad P'$,
Int$P'$ etc in the same way as it has been done in Remark
{}from \S\,2.4.

\smallskip

{\bf 2.} Up to now we have considered a complex gauge group $\bf G$.
The point is that in reduction of hCSW theories to 4D CFT theories
the imposing of reality conditions in some cases can be more tricky
than the choice of a compact subgroup $G\subset {\bf G}:=G^{\C}$ (see
e.g.~\cite{AW,AHS}). Later on we shall return to this problem.

\smallskip

{\bf 3.} From the discussions in \S\S\,3.2,3.3 one can see that the
description of holomorphic bundles is analogous to the description
of locally constant bundles. The difference is mainly in the
sheaves describing these theories.

\smallskip

\subsection{Theories on Calabi-Yau 3-folds }
\label{3.4}

Let us consider a CY 3-fold $\cz$. Then, besides a complex structure
$\cj$ on $\cz$, there exist a K\"ahler 2-form $\o$ and a Ricci-flat
metric, and the canonical line bundle of $\cz$ is trivial. There is
a one-to-one correspondence between complex structures on the bundle
$E' \to \cz$ and the operators $\bar\p_{\hat B}=\bar\p+\hat B$
satisfying the integrability conditions $\bar\p^2_{\hat B}=0$.
These operators do not depend on the K\"ahler form $\o$ and on the
(3,0)-form $\hat\Omega$ from (3.2) (a section of the canonical
 bundle of $\cz$). Hence the description of \S\S\,3.2,3.3 of the
moduli space of holomorphic bundles $E' \to \cz$ is not changed.

\smallskip

Some refinement arises if we impose an additional (mild) topological
condition of stability on the bundle $E'$ (for definitions,
 see~\cite{Do,UY} and references therein). Then the curvature 2-form
$F$ of a connection on $E'$ has to satisfy the additional equations
$$
F\wedge\o\wedge\o =0,
\eqno(3.14)
$$
which are equivalent to~\cite{Do,UY}
$$
\Lambda F=0.
\eqno(3.15a)
$$
Here $\Lambda$ is an algebraic `trace' operation inverse to
multiplication by $\o$. In local coordinates $\{z^a\}$ eqs.(3.15a)
have the form
$$
\o^{a\bar b}F_{a\bar b} = 0,
\eqno(3.15b)
$$
where $\{ \o^{a\bar b}\}$ are components of the bivector which is
inverse to the K\"ahler (1,1)-form
 $\o =\o_{a\bar b}dz^a\wedge dz^{\bar b}$.

\smallskip

Connections on the holomorphic bundle $E'$ satisfying eqs.(3.15)
are called the {\it Hermitian-Einstein} or {\it Hermitian-Yang-Mills}
 connections~\cite{Do,UY}. Solutions of eqs.(3.15) depend on
the K\"ahler 2-form $\o$. The moduli space $\cm^s_\cz$ of
Hermitian-Yang-Mills connections is a (dense) subset in the set
(3.12), $\cm^s_\cz \subset \cm_\cz$. Equations (3.6), (3.15) are
the field equations of the hCSW theory on a CY 3-fold $\cz$.

\smallskip

The study of holomorphic CSW theories on CY 3-folds $\cz$ is of
interest because of their
connection with $N$=2 topological strings~\cite{Wit2,BCOV}. These
theories are also connected with 4D CFT's, which we shall discuss
below in \S\,4. The study of geometry of the moduli space of stable
holomorphic bundles $E'$ over CY 3-folds is also important for
understanding  extended mirror symmetry and compactification of
the heterotic strings (see e.g.~\cite{Va} and references therein).
Extended mirror symmetry exchanges pairs $(\cz ,E')$ and
$(\tilde\cz ,\tilde E')$, where $\cz$,$\tilde\cz$ are CY 3-folds
and $E'\to \cz$,  $\tilde E'\to \tilde\cz$ are stable holomorphic
bundles. One of these bundles may coincide with the tangent bundle.
We recover standard mirror symmetry if both of these bundles are
tangent bundles. The discussion of this interesting topic is beyond
the scope of our paper.

\section{Integrable 4D conformal field theories}
\label{4}

\subsection{Holomorphic CSW theories on special complex 3-manifolds}
\label{4.1}

A choice of different gauge groups and coupling constants leads to
different CS/hCSW theories. As has been discussed in \S\,2.5, CS theories
on a real 3-manifold $X$ are equivalent to 2D CFT's if one supposes
that $X$ is a {\it trivial fibre bundle} over a 2-manifold $\Sigma$
with the fibre $\R$ or $S^1$. Analogously, holomorphic CSW theories
on a complex 3-manifold $\cz$ are connected with 4D CFT's if one
supposes that $\cz$ is a {\it bundle} over a real 4-manifold $M$ with
two-dimensional fibres.

\smallskip

We consider two main cases. In the first case a {\it complex}
3-manifold $\cz$ is the total space of a {\it fibre bundle}
over a real oriented 4-manifold $M$ with $S^2$ as a typical fibre
(Riemann surface of genus zero). Then, considering $\cz$ as a bundle
associated with the bundle of orthonormal frames on $M$, it may be
proved that the Weyl tensor of the manifold $M$ with a metric $g$ is
self-dual and such manifolds $M$ are called
{\it self-dual}~\cite{Pen,AHS},\cite{Wa2}-\cite{H1}. In the second case
we shall consider a {\it Calabi-Yau} 3-fold $\cz$ which is a
direct product of a 4-manifold $M$ and a torus $T^2$ (Riemann
surface of genus one) and therefore $\cz =M\times T^2$ is a
{\it trivial fibre bundle} over $M$. In this case $M$ must be a
hyperK\"ahler 4-manifold, for instance, $K3$ or $T^4$
with the changed canonical orientations.

\smallskip

Remember that the 2D WZNW model is a `generating' model for rational
2D CFT's. An analogous role in four dimensions is played by the
self-dual Yang-Mills (SDYM) model. The SDYM theory is connected
with the holomorphic CSW theory in the same way as the (chiral)
WZNW theory is connected with the ordinary CS theory. Indeed, in the
case when $\cz$ is a bundle over a self-dual 4-manifold $M$ (then
$\cz\equiv\cz (M)$ is the twistor space of $M$) the hCSW model
on $\cz$ can be {\it reduced} to the SDYM model on $M$. This follows
{}from the existence of a one-to-one correspondence between {\it self-dual}
vector bundles $E$ (bundles with self-dual connections $A$) over
$M$ and such holomorphic vector bundles $E'$ over the twistor
space $\cz (M)$ that are holomorphically trivial on any projective
line $\cp \hra \cz (M)$ in $\cz (M)$~\cite{Wa1}-\cite{ADHM}. From
the group-theoretic and cohomological point of view this
correspondence has been analyzed in~\cite{Po}.

\smallskip

We shall show that the SDYM model can also be obtained from the
hCSW model on a CY 3-fold $\cz =M\times T^2$. In this case gauge
{}fields `live' on a hyperK\"ahler 4-manifold $M$. Notice that
by using the so-called Yang gauge~\cite{Ya} the field equations
and action~\cite{Do,NS} for the SDYM model so resemble the field
equations and action for the 2D WZNW model that in the
paper~\cite{LMNS} the SDYM theory was called the 4D WZW theory.

\smallskip

In the case of the ordinary CS theory the moduli space of flat
connections on a bundle over $X = \Sigma \times \R$ is {\it bijective}
to the moduli space of flat connections on the induced bundle
over $\Sigma$, since the connection component along $\R$ can always be
removed by a gauge transformation. In the case of the holomorphic
CSW theory on a complex 3-manifold $\cz$, which is a fibration
over a 4-manifold $M$, for holomorphic connections it is impossible
to remove their components along fibres, since fibres are
two-dimensional as real manifolds. This is why the moduli space
$\cm _M$ of a 4D conformal field theory on $M$ connected with the hCSW
theory on $\cz$ is a {\it subspace} in the moduli space $\cm _\cz$ of the
hCSW theory (see \S\,3.3). Below in \S\,4.2 we shall describe an
embedding of the moduli space $\cm _M$ of self-dual gauge fields
on $M$ into the moduli space $\cm _\cz$ of flat (0,1)-connections on
the bundle $E'\to\cz$.

\subsection{Moduli space of self-dual Yang-Mills connections}
\label{4.2}

Consider a complex 3-manifold $\cz$ which is the fibre bundle with
the canonical projection $\pi :\cz\to M$
and fibres $\cp$ over  the points $x$ from a self-dual 4-manifold
 $M$. The typical fibre $\C\bp^1$ has the $SU(2)$-invariant
complex structure $\fj$ (see~\cite{Po}), and the vertical
 distribution
$V=\ker\pi_*$ inherits this complex structure. A restriction of $V$
to each fibre $\cp$, $x\in M$, is the tangent bundle to that fibre.
The Levi-Civita connection on the Riemannian manifold
$M$ generates the splitting of the
tangent bundle $T(\cz)$ into a direct sum
$$
T(\cz)=V\oplus H
\eqno(4.1)
$$
of the vertical $V$ and horizontal $H$ distributions.

\smallskip

Using the complex structures $\fj$ and $\cj$ on $\C\bp^1$
and $\cz$ respectively, one can split the complexified tangent bundle
of $\cz$ into a direct sum
$$
T^{\C}(\cz)=(V^{1,0}\oplus H^{1,0})\oplus (V^{0,1}\oplus H^{0,1})
\eqno(4.2)
$$
of subbundles of type (1,0) and (0,1). So we have the integrable
distribution $V^{0,1}$ of antiholomorphic vector fields.

\smallskip

Having the canonical distribution $V^{0,1}$ on the space $\cz$,
we introduce the sheaf $\cs$ of {\it partially holomorphic}
maps $\psi : \cz\to \bf G$, which are annihilated by vector
{}fields from $V^{0,1}$.  In other words, sections of the sheaf $\cs$
over open subsets $\U=U\times\Omega \subset\cz$ $\ (U\subset M,
\Omega\subset\C\bp^1)$ are $\bf G$-valued functions $\psi$
on $\U$, which are holomorphic on $\Omega\subset\cp\hra\cz$, $x\in U$.
It is obvious that the sheaf $\ch\equiv\hat\ch$ of holomorphic
maps from $\cz$ into $\bf G$, i.e. smooth maps which are
annihilated by vector fields from $V^{0,1}\oplus H^{0,1}$,
is a subsheaf of $\cs$, and $\cs$ is a subsheaf of the sheaf
$\hat\cs$ from \S\,3.2. Notice that in this section we use the
notation $\ch$ without hat for the sheaf $\hat\ch$.

\smallskip

Consider now the sheaves $\hat\cb^{0,q}$ introduced in \S\,3.2.
Let $\cb^{0,1}$ be the subsheaf of (0,1)-forms from $\hat\cb^{0,1}$
vanishing on the distribution $V^{0,1}$.
The sheaf $\cs$ acts on the sheaves $\cb^{0,1}$ and $\hat\cb^{0,q}$
by means of the adjoint representation. In particular, for $\cb^{0,1}$
and $\hat\cb^{0,2}$ we have the same formulae (3.3) with the replacement
$\hat\psi$ by $\psi\in \cs(\U)$,
$$
B\mapsto\mbox{Ad}_{\psi} B=
\psi^{-1}B\psi + \psi^{-1}\bar\p\psi ,
\eqno(4.3a)
$$
$$
\hat F\mapsto \mbox{Ad}_{\psi}\hat F=\psi^{-1}\hat F\psi ,
\eqno(4.3b)
$$
where $B\in\cb^{0,1}(\U)$, $\hat F\in \hat\cb^{0,2}(\U)$.

\smallskip

The map $\bar\d^0$, introduced in \S\,3.2, induces a map
$\bar\d^0: \cs\to\cb^{0,1}$, defined for any open set $\U$ of the
space $\cz$ by the formula
$$
\bar\d^0\psi = - (\bar\p\psi )\psi^{-1},
\eqno(4.4)
$$
where $\psi\in \cs(\U)$, $\bar\d^0\psi\in \cb^{0,1}(\U)$. Analogously,
the operator $\bar\d^1$ induces a map $\bar\d^1:  \cb^{0,1}\to
\hat\cb^{0,2}$, given for any open set $\U\subset\cz$ by the formula
$$
\bar\d^1 B=\bar\p B+B\wedge B,
\eqno(4.5)
$$
where $B\in \cb^{0,1}(\U)$,  $\bar\d^1 B\in\hat\cb^{0,2}(\U)$.

\smallskip

At last, let us denote by $\cb$ the subsheaf of $\cb^{0,1}$
consisting (of germs) of $\bf g$-valued (0,1)-forms
$B$ such that $\bar\d^1B=0$, i.e. sections $B$ of the sheaf
$\cb=\ker\bar\d^1$ satisfy the equations
$$
\bar\p B+B\wedge B=0.
\eqno(4.6)
$$
Restricting $\bar\d^0$  to $\cs\subset\hat\cs$
and $\bar\d^1$  to $\cb^{0,1}\subset\hat\cb^{0,1}$,
we obtain the exact sequence of sheaves
$$
{\bf 1}\lra\ch\stackrel{i}{\lra}\cs\stackrel{\bar\d^0}{\lra}\cb^{0,1}
\stackrel{\bar\d^1}{\lra}\hat\cb^{0,2},
\eqno(4.7)
$$
where ${\bf 1}$ is the identity of the sheaf $\ch$. From (4.7) we obtain
the exact sequence of sheaves
$$
{\bf 1}\lra\ch\stackrel{i}{\lra}\cs\stackrel{\bar\d^0}{\lra}
\cb\stackrel{\bar\d^1}{\lra}0,
\eqno(4.8)
$$
since $\im\bar\d^0=\ker\bar\d^1$ (the exactness of the sequence (4.7)),
and $\cs$ acts on $\cb$ transitively ($\cb\simeq \cs/\ch$).

\smallskip

{}From (4.8) we obtain the exact sequence of cohomology sets
$$
e\lra H^0(\cz, \ch)\stackrel{i_*}{\lra}H^0(\cz, \cs) \stackrel
{\bar\d^0_*}{\lra}H^0(\cz,\cb)\stackrel{\bar\d^1_*}{\lra}
H^1(\cz,\ch)\stackrel{\vp}{\lra} H^1(\cz, \cs) ,
\eqno(4.9)
$$
where the map $\vp$ is an embedding, induced by the
embedding of sheaves $i: \ch\to \cs$.  The kernel
$\ker\vp=\vp^{-1}(e)$ of the map $\vp$ coincides with a subset of
those elements from $H^1(\cz,\ch)$, which are mapped into the class
$e\in H^1(\cz,\cs)$ of smoothly trivial bundles over $\cz$, which are
holomorphically trivial on any projective line $\cp\hra\cz$, $x\in
M$. This means that representatives of the subset $\ker\vp$ of the
1-cohomology set $H^1(\cz,\ch)$ are those transition matrices $\cf\in
Z^1(\fu,\ch)$ for which there exists a decomposition
$$
\cf_{\a\b}=\psi_{\a}^{-1}\psi_{\b}
\eqno(4.10)
$$
with smooth $\bf G$-valued functions $\{\psi_{\a}\}\in C^1(\fu ,S)$ that
are {\it holomorphic} in local complex coordinate $\l$ on $\C\bp^1$.
Here $\fu=\{\U_{\a}\}$ is the cover of the space $\cz$ introduced
 already in \S\,3.1.

\smallskip

In the case under consideration it follows from the twistor
correspondence~\cite{Wa1}-\cite{AHS} that the moduli space
$\cm_M^{\C}$ of
complex self-dual gauge fields on a self-dual 4-manifold $M$ is
parametrized by the set $H^0(\cz,\cb)/H^0(\cz,\cs)$, where
$H^0(\cz,\cb)$ is the space of global solutions to eqs.(4.6), and
the group $H^0(\cz,\cs)$ is isomorphic to the group of gauge
transformations on the self-dual bundle $E$ over $M$~\cite{Po}. From
the exactness of the sequence (4.9) it follows that the set
$\ker \vp \subset H^1(X,\ch)$ is bijective to the moduli space
$\cm_M^{\C}$ of (complex) solutions to the SDYM equations on $M$,
$$
\cm_M^{\C}= H^0(\cz,\cb)/H^0(\cz,\cs)\simeq
\ker \vp \subset H^1(X,\ch).
\eqno(4.11)
$$
If we want to obtain real self-dual gauge fields with values in the
compact Lie algebra $\fg \subset \bf g$ and their moduli space
$\cm_M \subset \cm_M^{\C}$, we should impose some
specific reality conditions on all objects ({\bf G}-valued
{}functions $\psi_\a$, (0,1)-forms $B$ etc). For discussion of
the reality conditions see e.g.~\cite{AW}-\cite{ADHM}.

\subsection{Self-dual gauge fields on hyperK\"{a}hler manifolds}
\label{4.3}

In \S\,3.4 we discussed holomorphic CSW theories on a CY 3-fold
$\cz$. It was recalled that after imposing the stability condition,
connections on the holomorphic bundle $E'$ over $\cz$ have to
satisfy the Donaldson-Uhlenbeck-Yau equations (3.6),(3.15). To go
over to four dimensions, let us consider the simplest case
$\cz = M \times T^2$ (see \S\,4.1), where $M$ is a 4-manifold.
Then from the CY restriction it follows that $M$ is a hyperK\"ahler
manifold. Recall that hyperK\"ahler 4-manifolds are a particular
case of self-dual 4-manifolds, where in addition to self-duality
of the Weyl tensor the Ricci tensor is equals to zero~\cite{AHS,Be}.

\smallskip

So, we have the trivial bundle $\cz = M \times T^2 \to M$ with the
{}fibre $T^2$. Suppose, as in \S\,4.2, that components of a
connection on $E'$ along the fibre $T^2$ of the bundle
 $\cz \to M$ are equal to zero. Then the Donaldson-Uhlenbeck-Yau
equations on $\cz$ are reduced to the SDYM equations on the
 hyperK\"ahler manifold $M$. Thus, holomorphic CSW theories on CY
3-folds are also connected with SDYM theories but now defined on
hyperK\"ahler 4-manifolds. Clearly, the SDYM is a very distinguished
theory.

\smallskip

Besides the two analyzed cases (fibres $S^2$ and $T^2$ in the bundle
$\pi : \cz \to M$) it would be interesting to consider also the
{}following more general cases: (1) components of a connection on
the bundle $E' \to \cz$ are not equal to zero along fibres of the
bundle $\cz \to M$; (2) the space $\cz$ is a (nontrivial) bundle
over a 4-manifold $M$ with a Riemann surface $\Sigma$ of genus
$\ge 2$ as a typical fibre.

\subsection{Moduli space of local solutions to the SDYM equations}
\label{4.4}

It has been noted many times that
the chiral WZNW theory on a disk $D_0 \simeq \C{\bp}^1 -
\{\infty\}$ is one of the most important theories among 2D CFT's.
It is recovered from the CS theory
on $X = D_0 \times \R$~\cite{Wit1}-\cite{EMSS}. In four dimensions,
to this model there corresponds the SDYM model on an open set
$U \subset M$, which is in a sense `generic' since it describes all
{\it local} solutions of the SDYM equations. We shall briefly show how
to recover this model.

\smallskip

{}Fix an open set $U\subset M$ such that
$\cz|_U\simeq U\times \C\bp^1$  and choose coordinates $x^\mu$ on
$U$.  Consider the restriction of the twistor bundle $\pi :\cz\to M$
to $U$ and put $\rp :=\cz|_U$. The space $\rp$ is an open subset of
$\cz$, and, as a real manifold, $\rp$ is diffeomorphic to the direct
product $U\times\C\bp^1$. A metric $g$ on $U$ is not flat, and a
conformal structure $[g]$ on $U$ is coded into a complex structure
 $\cj$ on $\rp$~\cite{Pen, AHS}.  We again have a natural
one-to-one correspondence between solutions of the SDYM equations on
$U$ and holomorphic bundles $E'$ over $\rp$, holomorphically trivial
on (real) projective lines $\cp\hra\rp$, $\forall x\in U$.

\smallskip

Having described in \S\,4.2 the connection between the hCSW theory
on $\cz$ and the SDYM theory on $M$, we consider smooth solutions
$A$ of the SDYM equations on $U$ (local solutions). The moduli space
$\cm_U$ of real local solutions to the SDYM equations on $U$ is an
infinite-dimensional space and contains germs of all global
solutions on $M$. The moduli space $\cm_U$ and the moduli space
$\cm_U^{\C}$ of complex solutions have been described
in~\cite{Po}. In particular, it is shown that $\cm_U^{\C}$ is a
{\it double coset space}
$$
\cm_U^{\C}\simeq C^0(\fu, \ch )\bl C^1(\fu, \ch )/
C_{\t}^1(\fu, \ch ),
\eqno(4.12)
$$
where the groups of cochains $C^0(\fu, \ch ), C^1(\fu, \ch )$ of the
cover $\fu$ of the space $\rp$ with values in $\ch$ are considered as
{\it local groups}, and the group $C_{\t}^1(\fu, \ch )$ is a
 (diagonal) subgroup in $ C^1(\fu, \ch )$~\cite{Po}. The action
 $\r_0$ of the group $C^0(\fu, \ch )$ on the space $C^1(\fu, \ch
)/C_\t^1(\fu, \ch )$ is described in Appendix B. Analogously, $\cm_U$
is a {\it double coset space}
$$
\cm_U \simeq C_{\tau}^0(\fu, \ch )\bl C_{\tau}^1(\fu, \ch )/
C_{\tau\t}^1(\fu, \ch ),
\eqno(4.13)
$$
where the real subgroups $C_{\tau}^0(\fu, \ch )$, $C_{\tau}^1(\fu,
\ch )$ and $C_{\tau\t}^1(\fu, \ch )$ of the local groups
$C^0(\fu, \ch )$, $C^1(\fu, \ch )$ and $C_{\t}^1(\fu, \ch )$
are described in~\cite{Po}. Concerning (4.12) and (4.13), recall
that $K\bl G/H$ denotes the double coset space resulting from
division of a group $G$ by the right action of $H$ and left action
of $K$.

\smallskip

To sum up, self-duality in 4D is connected with holomorphy in 6D
and is a generalization of chirality in 2D. The SDYM model on an
open ball $U \subset \R^4$ is an analogue of the chiral WZNW
model on a disk $D_0 \subset \R^2 \simeq \C$. The {\it double
coset structure} (4.13) of the moduli space of solutions to the
SDYM equations on $U$ is an analogue of the {\it coset structure}
$LG/G$ of the moduli space of solutions to the chiral WZNW model
on $D_0$. So, quantization of the SDYM model on an open ball
$U \subset \R^4$ is an urgent task for constructing of integrable
4D conformal quantum field theories.

\subsection{Relatives of the SDYM model}
\label{4.5}

Having considered the basic nonlinear integrable 4D conformal field
theory - the SDYM theory - one may move on to a consideration of its
``relatives". Just as for ordinary CS theories, it is interesting
to study holomorphic CSW theories on the twistor space $\cz (M)$
of a self-dual 4-manifold $M$ when we have:

1) connected and simply connected gauge group $\bf G$,

2) connected but not simply connected gauge group $\bf G$,

3) disconnected groups of type $N\ltimes \bf G$, where $\bf G$ is a

connected group with a discrete automorphism group $N$.

\smallskip

The case 1) is standard. The case 2) was also considered in the
literature, where it was shown, for instance, that instantons for
non-simply connected gauge groups can have fractional topological
charge (see e.g.~\cite{Ho}). It seems that the case 3) of
disconnected groups was not discussed in the literature on the SDYM
model. In this case we can obtain analogues of {\it orbifold
models} studied in 2D CFT's~\cite{DVV}.

\smallskip

It also seems interesting to study {\it coset} SDYM models
which can be introduced analogously to 2D/3D coset models
(see e.g.~\cite{MS}) by using the Donaldson-Nair-Schiff
 action~\cite{Do,NS} and appropriate boundary conditions. It is
strange that such models have not been considered in the literature.

\newpage
\section{Solvable theories in two, four and six dimensions}
\label{5}

\subsection{Free conformal field theories in four dimensions}
\label{5.1}

Among 2D quantum field theories the important role is played by
{}free conformal field theories (see e.g.~\cite{FMS} and references
therein). Among free 4D CFT's on a 4-manifold $M$ there is an
important subclass of completely solvable models connected with
the holomorphic CSW theory on the twistor space $\cz (M)$ of $M$.
We shall first describe these theories in 4D and then
in 6D setting.

\smallskip

Let us consider a self-dual 4-manifold $M$ with a metric $g$. Let
we are given a homeomorphism of an open subset $U \subset M$ on
an open subset of the space $\R^4$. Then the coordinates $x^{\mu}$
on $\R^4$, $\mu , \nu ,...= 1,...,4$, define local coordinates
on $U \subset M$, which we also denote by $x^{\mu}$. Speaking
{}further about objects in the local coordinates, we  mean
$x^{\mu}$, e.g. for the metric $g$ in the local coordinates we have
$g = g_{\mu \nu}dx^{\mu}dx^{\nu}$.

\smallskip

On tangent spaces $T_{x}M$ we use the metric with the components
$\d_{\a \b}$, where $\a ,\b ,...=1,...,4$ are tangent indices, and
introduce a (local) orthonormal frame field
$e_\a =e_\a^\mu \p_\mu = e_\a^\mu \p /\p{x^\mu}$ for the tangent
bundle $TM$ and a (local) orthonormal coframe field
$\epsilon^\a = \epsilon^\a_\mu dx^\mu$ for the cotangent bundle $T^*M$,
$g = \d_{\a \b}\epsilon^\a \epsilon^\b$, $e_\a^\mu \epsilon^\b_\mu =
\d_\a^\b$.

\smallskip

We denote the Levi-Civita connection on $M$ by $\nabla$. Let
$\nabla_\a := \nabla_{e_\a}$ be its component along $e_\a$.
Using the Pauli matrices $\sigma^a=(\sigma^a_{AA'})$, where
$a, b, ...=1,2,3,\ A=1,2,\ A'=1,2$, we introduce the matrices
$\sigma^\a=\{i\sigma^a, 1\}$ with the components $\sigma^\a_{AA'}$
and put $\nabla_{AA'}:=\sigma^\a_{AA'}\nabla_\a$. Having raised the
index with the help of $\ve^{AB}=-\ve^{BA}$, $\ve^{12}=1$, we obtain
the operator $\nabla^A_{A'}:=\ve^{BA}\nabla_{BA'}$.

\smallskip

Let $\vp_{A_1...A_n}$ be a smooth complex n-index symmetric spinor
{}field on $M$ ($n>0$) with conformal weight $-1$. These fields have
spin $s=n/2$ and  negative helicity~\cite{PW, PenR}. They are
chiral primary fields in the terminology of 4D CFT's. Equations
of motion of such free massless fields interacting with the gravity
{}field (metric $g$ on $M$) have the following form~\cite{PW, PenR}:
$$
\nabla^{A_1}_{A'}\vp_{A_1A_2...A_n}=0.
\eqno(5.1a)
$$
One can also consider a scalar field $\vp_0$ of spin $s=0$, the field
equation for which is
$$
\square\vp_0 + \frac{1}{6}R\vp_0=0.
\eqno(5.1b)
$$
Here $\square :=\d^{\a\b}\nabla_\a\nabla_\b$ is the Laplacian on $M$, and
$R$ is the scalar curvature.

\smallskip

Equations (5.1) are  field equations of free 4D CFT's describing
chiral spinor fields of spin $s=n/2$ ($n\ge 0$) interacting
with a self-dual background gravitational  field $g$ (the Weyl
tensor for $g$ is self-dual). One can also introduce the
interaction of these spinor fields with self-dual Yang-Mills fields.
To do this, let us consider a principal $\bf G$-bundle $P\to M$
with a self-dual connection $D$ on $P$. Let $D_\a :=D_{e_\a}$ be its
component along $e_\a$. Denote by $\tilde\nabla_\a :=\nabla_\a\otimes 1
+ 1\otimes D_\a $ the components of tensor-product connection
containing both the Levi-Civita and Yang-Mills connections and introduce
the operator $\tilde\nabla^A_{A'}:=\ve^{BA}\sigma^\a_{BA'}
\tilde\nabla_\a$.

\smallskip

Denote by $\ad P$ a vector bundle with the base $M$ and a typical fibre
$\bf g$
generated by the adjoint representation of the group $\bf G$, $\ad P=
P\times_{\ad \bf G}\bf g$. Let $\tilde\vp_{A_1...A_n}$ be a smooth
n-index symmetric spinor field on $M$ ($n\ge 0$) with values in $\ad P$.
Locally this is a $\bf g$-valued field on open subsets of $M$.
Now one can introduce equations of motion of free massless chiral
fields $\tilde\vp_{A_1...A_n}$ interacting with a background self-dual
gauge and gravitational fields as~\cite{Hi2}
$$
\tilde\nabla^{A_1}_{A'}\tilde\vp_{A_1A_2...A_n} = 0,
\eqno(5.2a)
$$
$$
\tilde\square\tilde\vp + \frac{1}{6}R\tilde\vp =0,
\eqno(5.2b)
$$
where $\tilde\square := \d^{\a\b}\tilde\nabla_\a\tilde\nabla_\b$.
Equations (5.2) are field equations of 4D CFT's describing free
$\bf g$-valued chiral spinor fields of  spin $s=n/2$
($n\ge 0$) in a self-dual background.

\subsection{Free 4D conformal field theories in holomorphic setting}
\label{5.2}

{}Free 4D CFT's described in \S\,5.1 can be reformulated as
theories of free holomorphic fields on the twistor space $\cz$
of the self-dual manifold $M$~\cite{PW, PenR, Hi2}. Namely,
let us define the tautological line bundle $L\to\cz$ by taking
the restriction  $L_{|\cp}$ of $L$ to each fibre $\cp\hra\cz$
to be
the standard holomorphic tautological line bundle $L_x$ over
$\cp$ with the first Chern class $c_1(L_x)=-1$.
Denote by $L^k$ the $k$-th tensor power of the bundle $L$ and by
$\co (-k)$ the sheaf of holomorphic sections of the bundle
$L^k\to \cz$. Then the cohomology group $H^1(\cz , \co (-n-2))$
is isomorphic to the space $\cm_n$ of smooth solutions to
eqs.(5.1) on $M$~\cite{PW, PenR},
$$
\cm_n\simeq H^1(\cz , \co (-n-2)).
\eqno(5.3)
$$
Here $n\ge 0$ is the spin of the chiral spinor fields $\vp_{A_1...A_n}$
{}from (5.1).

\smallskip

Now let us consider fields $\tilde\vp_{A_1...A_n}$ on $M$
satisfying eqs.(5.2). Using the projection $\pi :\cz\to M$ of the
twistor space on the self-dual 4-manifold $M$ and the twistor
correspondence, we can pull back
the bundle $P\to M$ to the bundle $P':=\pi^*P$ over $\cz$. Let us
consider the associated bundles $\ad P':=P'\times_{\ad{\bf G}}\bf g$
and $\ad P'\otimes L^k$. A self-dual connection $D$ on the bundle $P$
induces the holomorphic connection on the bundle $\ad P'$ satisfying
eqs.(4.6) of the holomorphic Chern-Simons-Witten theory and
therefore the bundle $\ad P'\otimes L^k$ is holomorphic.
Denote by $\co^{\bf g}(-k)$ the sheaf of holomorphic sections of the
bundle $\ad P'\otimes L^k$. There is an isomorphism
$$
\cm_n^{\bf g}\simeq H^1(\cz , \co^{\bf g}(-n-2))
\eqno(5.4)
$$
between the first cohomology group $H^1(\cz , \co^{\bf g}(-n-2))$
of the space $\cz$ with the coefficients in the sheaf
$\co^{\bf g}(-n-2)$
and the solution space $\cm^{\bf g}_n$ of eqs.(5.2)~\cite{Hi2}.

\smallskip

Using the isomorphisms (5.3) and (5.4), one can write down formulae for
general solutions of eqs.(5.1) and (5.2). Thus, the above-described free
4D CFT's are not only integrable but also explicitly solvable.
However, we shall call them free integrable 4D CFT's.

\subsection{Interconnections of 4D and 6D theories}
\label{5.3}

In \S\S\,4,5.1,5.2 we have considered interconnections of
holomorphic Chern-Simons-Witten theories in three complex dimensions
(six real dimensions) with integrable conformal field theories in four
real dimensions. This consideration can be summarized in the following
``commutative diagram":
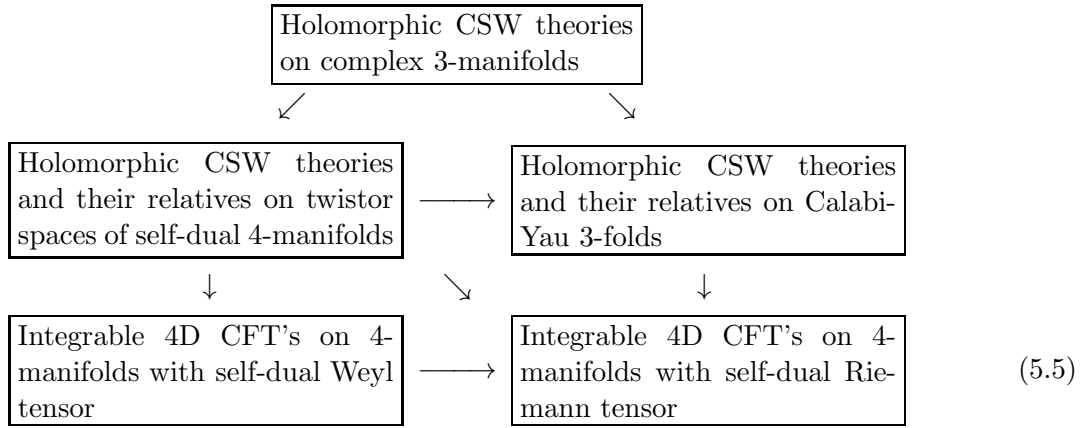
$$
\begin{CD}
@. \fbox{\parbox{4.7cm}{Holomorphic CSW theories on complex 3-manifolds}}@.\\
@.\swarrow\hspace{4cm}\searrow@. \\
\end{CD}
$$
$$
\begin{CD}
\fbox{\parbox{5cm}{Holomorphic CSW theories
and their relatives on twistor spaces of self-dual 4-manifolds}}@>>>
\fbox{\parbox{5cm}{Holomorphic CSW theories and their relatives on
Calabi-Yau 3-folds}}
\end{CD}
$$
$$
\downarrow\hspace{3cm}\searrow\hspace{3cm}\downarrow
$$
$$
\begin{CD}
\fbox{\parbox{5cm}{Integrable 4D CFT's on 4-manifolds with self-dual
Weyl tensor}}@>>>
\fbox{\parbox{5cm}{Integrable 4D CFT's on 4-manifolds with self-dual
Riemann tensor}}
\end{CD}
\eqno(5.5)
$$

\bigskip

\noindent
The arrows mean that one theory can be derived from another one.
{}For instance, integrable 4D CFT's on 4-manifolds with self-dual
Riemann tensor can be derived from hCSW theories on twistor spaces
and from hCSW theories on CY 3-folds.

\smallskip

Almost all theories described above arise in string theory. Namely,

\begin{itemize}
\item integrable 4D CFT's on 4-manifolds with the self-dual Riemann
      tensor are connected with $N=2$ strings
      (4D target space)~\cite{OV1,OV3};
\item holomorphic CSW theories on Calabi-Yau 3-folds are connected
      with $N=2$ topological strings
      (6D target space)~\cite{Wit2, BCOV};
\item holomorphic CSW theories on twistor spaces of Ricci-flat
      self-dual 4-manifolds  are connected with $N=4$ topological
      strings~\cite{BV,OV2};
\item holomorphic CSW theories on complex 3-manifolds can be
      connected with
      generalized $N=2$ topological string
      theories~\cite{BCOV, OV2}.
\end{itemize}
\noindent
It would be interesting to study the last case. Notice also that
heterotic $N= (2,1)$ strings~\cite{OV3,KM} are connected with
integrable 4D CFT's on 4-manifolds with the self-dual Weyl tensor,
and we shall describe this elsewhere.

\subsection{Parallels between 2D and 4D conformal theories}
\label{5.4}

In \S\,2.5 we have discussed 2D CFT's arising from ordinary CS
theories in 3 real dimensions, and in \S\,4 we have discussed 4D
CFT's arising from holomorphic CSW theories in 3 complex dimensions.
Parallels between 2D and 4D conformal field theories are summarized
in the following table:

\medskip
\begin{center}
\begin{tabular}{|c|c|}
\hline
2D & 4D\\[1mm]
\hline
chiral WZNW models & SDYM models\\[5mm]
\parbox{4.5cm}{WZNW models with non-simply connected gauge groups} &
\parbox{7cm}{SDYM models with non-simply connected gauge groups
describe instan\-tons with fractional topological charge}\\[1cm]
orbifold models & orbifold models (can be introduced)\\[5mm]
coset models   & coset models (can be introduced)\\[5mm]
{}free chiral CFT's &  \parbox{6cm}{free chiral CFT's describing (chiral)
{}fields of negative helicity and arbitrary spin $s\ge 0$} \\[5mm]
\hline
\end{tabular}
\end{center}

\medskip

In this paper we discussed how holomorphic CSW theories on complex
3-manifolds are connected with integrable CFT's on real 4-manifolds
and showed that these theories are similar to (chiral) rational 2D
CFT's or (chiral) free 2D CFT's. It is well known that significant
progress in understanding CFT's in two dimensions was related to
the existence of the Virasoro-Ka\v{c}-Moody symmetries imposing
severe restrictions on correlation functions and uniquely determining
them in genus zero. Naturally, in studing  integrable 4D CFT's the
{}following questions arise:

\smallskip

1. What are analogues of affine Lie algebras of 2D CFT's ?

2. What is an analogue of the Virasoro algebra ?

\smallskip

\noindent
We shall answer these questions in \S\,6 using the \v{C}ech approach
to the sheaf cohomologies. To see that symmetry algebras of
integrable 4D CFT's really generalize symmetry algebras of 2D
CFT's, in \S\,6.1 we shall describe
the Virasoro and affine Lie algebras
in terms of the \v{C}ech cohomology.

\newpage

\section{Cohomological symmetry algebras}
\label{6}

\subsection{The \v{C}ech description of the Virasoro and affine Lie algebras}
\label{6.1}

Let us consider a two-dimensional sphere $S^2$. The sphere $S^2$
can be covered by two coordinate patches $\Omega_1,\Omega_2$, with
$\Omega_1$ the neighbourhood of $\l =0$, and $\Omega_2$, the
neighbourhood of $\l =\infty$, where $\l$ is a complex coordinate
on $S^2 = \C \cup \infty$. The sphere $S^2$, considered as a complex
projective line $\C\bp^1 = \Omega_1 \cup \Omega_2$, is the complex
manifold obtained by patching together $\Omega_1$ and $\Omega_2$ with
the coordinates $\l$ on $\Omega_1$ and $\zeta$ on $\Omega_2$ related
by $\zeta =\l^{-1}$ on  $\Omega_1 \cap \Omega_2$. For example, if
$\Omega_1 =\{\l\in\C : |\l |<\infty\}$ and
$\Omega_2 =\{\l\in\C\cup\infty : |\l |>0\}$, $\Omega_1 \cap \Omega_2$
is the multiplicative group $\C^*$ of complex numbers $\l\ne\{0,\infty\}$.
In the following we shall consider this two-set open cover $\fo =
\{\Omega_1 , \Omega_2\}$ of the Riemann sphere $\C\bp^1$.

\smallskip

Remember that the affine Lie algebra ${\bf g}\otimes\C[\l ,\l^{-1}]$ (without
a central term) is the algebra of ${\bf g}$-valued meromorphic functions
on $\C\bp^1\simeq\C^*\cup\{0\}\cup\{\infty\}$ with the poles at
$\l =0, \l =\infty$
and holomorphic on $\Omega_{12}=\Omega_1\cap\Omega_2\simeq\C^*$. Hence, it
is a subalgebra in the algebra
$$
C^1(\fo , \co^{\bf g}_{\C\bp^1})\simeq {\bf g}\otimes\C[\l ,\l^{-1}]\oplus
{\bf g}\otimes\C[\l ,\l^{-1}]
\eqno(6.1)
$$
of 1-cochains of the cover $\fo =\{\Omega_1,\Omega_2\}$ of  $\C\bp^1$
with values in the sheaf of holomorphic maps from $\C\bp^1$
into the Lie algebra ${\bf g}$. For definitions, see
Appendices. Notice that (central) extensions of the
algebra (6.1) will appear after passing to quantum theory.

\smallskip

Elements of the Virasoro algebra $Vir^0$ (with zero central charge) are
meromorphic vector fields on $\C\bp^1$ having poles at the points
$\l =0, \l =\infty$ and holomorphic on the overlap $\Omega_{12}=\Omega_1
\cap\Omega_2\simeq\C^*=\C\bp^1-\{0\}-\{\infty\}$. This algebra has the
{}following \v{C}ech description. Let us consider the sheaf $\cv_{\C\bp^1}$
of holomorphic vector fields  on $\C\bp^1$. Then, for the space of \v{C}ech
1-cochains with values in $\cv_{\C\bp^1}$ we have
$$
C^1(\fo , \cv_{\C\bp^1})\simeq  Vir^0\oplus Vir^0.
\eqno(6.2)
$$
Notice that for $\{v_{12},v_{21}\}\in C^1(\fo , \cv_{\C\bp^1})$ the
antisymmetry condition cannot be imposed on cohomology indices of the
holomorphic vector fields $v_{12},v_{21}$ since it is not preserved
under commutation. So in the general case we have $v_{21}\ne -v_{12}$.

\smallskip

The space  $Z^1(\fo , \cv_{\C\bp^1})=\{v\in C^1(\fo , \cv_{\C\bp^1}):
v_{12}=-v_{21}\}$ of 1-cocycles of the cover
$\fo$ of  $\C\bp^1$ with values in the sheaf
$\cv_{\C\bp^1}$ coincides with the algebra $Vir^0$ as a vector space
since
$$
Z^1(\fo , \cv_{\C\bp^1})\simeq
(Vir^0\oplus Vir^0)/diag(Vir^0\oplus Vir^0).
\eqno(6.3)
$$
{}Further, by virtue of the equality
$$
H^1(\C\bp^1, \cv_{\C\bp^1})=0,
\eqno(6.4)
$$
which means the rigidity of the complex structure of $\C\bp^1$,
any element $v_{12}=-v_{21}$ from $Z^1\simeq Vir^0$ can be represented
in the form
$$
v_{12}=v_1-v_2.
\eqno(6.5)
$$
Here, $v_1$ can be extended to a holomorphic vector field on $\Omega_1$,
and $v_2$ can be extended to a holomorphic vector field on $\Omega_2$.

\smallskip

It follows from (6.3)--(6.5) that the algebra $Vir^0$ is connected
with the algebra
$$
C^0(\fo , \cv_{\C\bp^1})
\eqno(6.6)
$$
of 0-cochains of the cover $\fo$ with values in the sheaf $\cv_{\C\bp^1}$
by the (twisted) homomorphism
$$
{}f^0: C^0(\fo , \cv_{\C\bp^1})\lra C^1(\fo , \cv_{\C\bp^1})
\Leftrightarrow
\eqno(6.7a)
$$
$$
{}f^0: \{v_1,v_2\}\mapsto \{v_1-v_2,v_2-v_1\}.
\eqno(6.7b)
$$
Just the cohomological nature of the algebra $Vir^0$ permits one to
define its local action on Riemann surfaces of an arbitrary genus and on
the space of conformal structures of Riemann surfaces~\cite{BMS}. A
central extension arises under an action of the Virasoro algebra on
holomorphic sections of line bundles over moduli spaces
(quantization).

\subsection{The twistor space $\rp$: some definitions}
\label{6.2}

If the Virasoro and affine Lie algebras of symmetries
arise while considering 2D CFT's on a disk $D_0\simeq
\C\subset\C\bp^1$, as described in \S\,6.1,  their 4D
analogues arise as symmetry algebras of integrable 4D
CFT's on an open ball $U\subset\R^4$. In particular, these
symmetry algebras appear in the SDYM theory on $U\subset\R^4$
describing a subsector of the hCSW theory on the twistor
space $\rp$  of $U$~\cite{Po}.

\smallskip

As a smooth 6D manifold the
twistor space ${\rp}\equiv{\rp}(U)$ of $U$ is a direct product
of the spaces $U$ and $\C{\bp}^1$: ${\rp}=U\times\C{\bp}^1$
and is the bundle of complex structures on $U$~\cite{AHS}. This space
can be covered by two coordinate patches,
$$
{\rp}=\U_1\cup\U_2,\ \U_1=U\times\Omega_1,\ \U_2=U\times\Omega_2,
\eqno(6.8)
$$
with the coordinates $\{x^\mu, \l ,\bar\l \}$ on $\U_1$ and $\{x^\mu,
\zeta , \bar\zeta \}$ on $\U_2$. The two-set open cover $\fo =\{\Omega_1,
\Omega_2\}$ of the Riemann sphere $\C{\bp}^1$ has been described in
\S\,6.1.  We shall consider the intersection $\U_{12}$ of $\U_1$
and $\U_2$,
$$
\U_{12}:=\U_1\cap\U_2=U\times(\Omega_1\cap\Omega_2),
\eqno(6.9)
$$
with the coordinates $x^\mu$ on $U$, $\l ,\bar\l$ on $\Omega_{12}:=
\Omega_1\cap\Omega_2$. Thus, the twistor space $\rp$ is a trivial bundle
$\pi : {\rp}\to U$ over $U$ with the fibre $\C{\bp}^1$, where
$\pi : \{x^\mu , \l , \bar\l \}\to \{x^\mu\}$ is the canonical
projection. For more details see~\cite{Po}.

\smallskip

The space $\rp$ is a complex manifold. One can introduce complex
coordinates $\{z^a_1\}$ on $\U_1$,  $\{z^a_2\}$ on $\U_2$, and
on the intersection $\U_{12}$ of charts $\U_1$ and $\U_2$ these
coordinates are connected by the holomorphic transition function
$f_{12}$, $z^a_1= f^a_{12}(z^b_2)$ (see~\cite{Po} for explicit
expressions).

\smallskip

In \S\S\,6.3, 6.4 we shall briefly describe symmetry algebras of
the SDYM equations on $U$ answering thus the questions of \S\,5.4
on 4D analogues of the Virasoro and affine Lie algebras (without
central terms).

\subsection{Analogues of affine Lie algebras in integrable 4D CFT's}
\label{6.3}

An affine-type symmetry algebra of integrable 4D CFT's is connected
with the algebra $\cg_h$ of functions that are holomorphic on
$\U_{12}=\U_1\cap\U_2\subset \rp$ and take values in the Lie algebra
$\bf g$ of a complex Lie group $\bf G$.  The algebra $\cg_h$ with
pointwise commutators generalizes affine Lie algebras. The symmetry
algebra is the algebra $$ C^1(\fu ,\co^{\bf g}_\rp )\simeq
\cg_h\oplus \cg_h
\eqno(6.10)
$$ of 1-cochains of the cover $\fu
=\{\U_1,\U_2\}$ of the space $\rp$ with values in the sheaf $\co^{\bf
g}_\rp$ of holomorphic maps from $\rp$ into the Lie algebra ${\bf
g}$. We mainly consider the case ${\bf g} =sl(n,\C)$.

\smallskip

The algebra (6.10) acts on the transition matrix $\cf^0_{12}=1$
of the {\it trivial} holomorphic bundle $E_0'=\rp\times\C^n$
corresponding to the vacuum solution $A^0=0$ of the SDYM equations
on $U$ in the Penrose-Ward construction. This action generates
infinitesimal deformations of the bundle $E_0'$ described by $\d\cf^0$
and variations $\d A^0$ of the trivial solution  $A^0$.
If we want to consider infinitesimal variations of a nontrivial
self-dual gauge potential $A$ and a transition matrix $\cf_{12}$
in a holomorphically nontrivial bundle $E'\to\rp$,  we have to
use a  more general sheaf than the sheaf $\co^{\bf g}_\rp$ from
(6.10)~\cite{ita}.
Namely, consider a principal fibre bundle $P'\to\rp$ with a transition
matrix $\cf_{12}$ on $\U_{12}=\U_1\cap\U_2$ for the cover $\fu =
\{\U_1, \U_2\}$ of $\rp$.
Then, introduce the associated bundle $\ad P'=P'\times_{\ad\bf G}\bf g$
of Lie algebras and the sheaf of holomorphic sections of the
bundle $\ad P'$. Since this sheaf is isomorphic to the sheaf
$\co^{\bf g}_\rp$  considered above, we also denote it by
$\co^{\bf g}_\rp$.

\smallskip

Now we introduce the algebra of 1-cochains of the cover $\fu$ with
values in the sheaf of holomorphic sections of the bundle $\ad P'$.
This algebra is isomorphic to the algebra (6.10), and we also
denote it by $C^1(\fu , \co^{\bf g}_\rp )$. For a description of
infinitesimal symmetries of the SDYM equations we need also the sheaf
$\cs^{\bf g}_\rp$ of {\it smooth} maps from $\rp$ into $\bf g$
which are holomorphic along fibres $\cp$ of the bundle $\rp\to U$.
Then we introduce the algebra $C^0(\fu , \cs^{\bf g}_\rp )$ of
0-cochains of the cover $\fu$ with values in $\cs^{\bf g}_\rp$.
Elements of the algebra $C^0(\fu , \cs^{\bf g}_\rp )$ are the collection
$\{\phi_1,\phi_2\}\in C^0(\fu , \cs^{\bf g}_\rp )$ of sections of
the sheaf $\cs^{\bf g}_\rp$, where $\phi_1$ is a section over
$\U_1$, and  $\phi_2$ is a section over $\U_2$. Now we briefly
describe the action of  $C^1(\fu , \co^{\bf g}_\rp )$ on $\cf_{12}$
and $A=A_{\mu}dx^\mu$ following the papers~\cite{Po, ita}.

\smallskip

The action of $\theta =\{\theta_{12}, \theta_{21}\}\in
C^1(\fu , \co^{\bf g}_\rp )$ on the transition matrix $\cf_{12}$ is
$$
\cf_{12}\mapsto\d_\theta\cf_{12}=\theta_{12}\cf_{12}-\cf_{12}\theta_{21} .
\eqno(6.11)
$$
If we choose the compact gauge Lie algebra $\fg=su(n)\subset {\bf g}=
sl(n,\C)$ and want to preserve the reality of gauge fields, then in
$C^1(\fu , \co^{\bf g}_\rp )$ we should choose a subalgebra (real form)
$C^1_{\tau}(\fu , \co^{\bf g}_\rp )$ imposing the conditions~\cite{Po,ita}
$$
\theta^{\dagger}_{12}(x, -\bar\l^{-1}) = -\theta_{21}(x,\l ),
$$
where the coordinates $x,\l ,\bar\l$ have been introduced in (6.8), and
$^\dagger$ denotes Hermitian conjugation.

\smallskip

Recall that to a self-dual gauge potential $A=A_\mu dx^\mu$ there
corresponds the (0,1)-part $\bar\p_B$ of a connection on the bundle
$E'\to\rp$ satisfying eqs.(4.6) and the transition matrix
$\cf_{12}=\psi^{-1}_1\psi_2$,
where $\bf G$-valued functions $\psi_{1,2}$ are defined from
the equations $(\bar\p +B)\psi_1=0$ and $(\bar\p +B)\psi_2=0$
on $\U_1$ and $\U_2$, respectively~\cite{Po}. We introduce $\psi_1
(\d_\theta\cf_{12})\psi^{-1}_2$ and from the formula
$\d_\theta\cf_{12}=(\d_\theta\psi^{-1}_1)\psi_2 +
\psi^{-1}_1\d_\theta\psi_2$ it follows that
$$
\psi_1 (\d_\theta\cf_{12})\psi^{-1}_2\in Z^1(\fu , \cs^{\bf g}_\rp ).
$$
Since $H^1(\C\bp^1,\co^{\bf g}_{\C\bp^1})=H^1(\rp ,\cs^{\bf g}_{\rp})=0$,
there always exists a splitting
$$
\psi_1 (\d_\theta\cf_{12})\psi^{-1}_2=\phi_1(\theta )-
\phi_2(\theta ),
\eqno(6.12)
$$
where $\{\phi_1(\theta ), \phi_2(\theta )\}\in C^0(\fu ,
\cs^{\bf g}_\rp )$. Smooth  $\bf g$-valued function $\phi_1(\theta )$
on $\U_1$, holomorphic in $\l$,  and smooth $\bf g$-valued function
$\phi_2(\theta )$ on $\U_2$, holomorphic in $1/\l$, give a
solution of the infinitesimal variant of the Riemann-Hilbert problem.
Moreover, formula (6.12) defines a (twisted) {\it isomorphism} between
the algebra $C^1(\fu , \co^{\bf g}_\rp )$ and a subalgebra in
$C^0(\fu , \cs^{\bf g}_\rp )$ obtained through solutions of
infinitesimal Riemann-Hilbert problems.

\smallskip

Having $\phi (\theta )=\{\phi_1(\theta ), \phi_2(\theta )\}\in C^0(\fu ,
\cs^{\bf g}_\rp )$, one can define an action of the algebra
$C^1(\fu , \co^{\bf g}_\rp )$ on smooth objects $\psi=\{\psi_1,
\psi_2\}$ and $B=\{B^{(1)}, B^{(2)}\}$ on $\rp$. Namely, this algebra
acts on $\psi_1, \psi_2$ and $B^{(1)}, B^{(2)}$ as follows
$$
\d_\theta\psi_1 = - \phi_1(\theta )\psi_1,\quad
\d_\theta\psi_2 = - \phi_2(\theta )\psi_2,
\eqno(6.13a)
$$
$$
\d_\theta B^{(1)} =\bar\p\phi_1(\theta ) + [B^{(1)},\phi_1(\theta )],\
\d_\theta B^{(2)} =\bar\p\phi_2(\theta ) + [B^{(2)},\phi_2(\theta )],
\eqno(6.13b)
$$
where $\phi (\theta )$ corresponds to $\theta$ through eqs.(6.12).
Notice that this is not a gauge transformation, because
$\phi_1(\theta )\ne \phi_2(\theta )$ on $\U_{12}\subset\rp$.

\smallskip

At last, one can define an action of the algebra $C^1(\fu ,
\co^{\bf g}_\rp )$  on the self-dual gauge potential $A=A_\mu dx^\mu$
using the complex coordinates $y^1=x^1+ix^2, y^2=x^3-ix^4,
\bar y^1=x^1-ix^2, \bar y^2= x^3+ix^4$ on $\R^4\simeq\C^2$. Let us
rewrite components of the connection on $E\to U$ in these coordinates:
$D_{y^1}=\p_{y^1} + A_{y^1},\ D_{y^2}=\p_{y^2} + A_{y^2},\
D_{\bar y^1}=\p_{\bar y^1} + A_{\bar y^1},\ D_{\bar y^2}=\p_{\bar y^2}
+ A_{\bar y^2}$. Then we have~\cite{Po, ita}
$$
\d_\theta A_{y^1}= \mathop{Res}\limits_{\l =0}[\l^{-2}
(D_{\bar y^2}\phi_2(\theta ) + \l D_{y^1}\phi_2(\theta ))]:=
\oint_{S^1}\frac{d\l}{2\pi i\l^2}
(D_{\bar y^2}\phi_2(\theta ) + \l D_{y^1}\phi_2(\theta )),
\eqno(6.14a)
$$
$$
\d_\theta A_{y^2}=- \mathop{Res}\limits_{\l =0}[\l^{-2}
(D_{\bar y^1}\phi_2(\theta ) - \l D_{ y^2}\phi_2(\theta ))]:=
-\oint_{S^1}\frac{d\l}{2\pi i\l^2}
(D_{\bar y^1}\phi_2(\theta ) - \l D_{ y^2}\phi_2(\theta )),
\eqno(6.14b)
$$
$$
\d_\theta A_{\bar y^1}= \mathop{Res}\limits_{\l =0}[\l^{-1}
(D_{\bar y^1}\phi_1(\theta ) - \l D_{y^2}\phi_1(\theta ))]:=
\oint_{S^1}\frac{d\l}{2\pi i\l}
(D_{\bar y^1}\phi_1(\theta ) - \l D_{y^2}\phi_1(\theta )),
\eqno(6.14c)
$$
$$
\d_\theta A_{\bar y^2}= \mathop{Res}\limits_{\l =0}[\l^{-1}
(D_{\bar y^2}\phi_1(\theta ) + \l D_{y^1}\phi_1(\theta ))]:=
\oint_{S^1}\frac{d\l}{2\pi i\l}
(D_{\bar y^2}\phi_1(\theta ) + \l D_{y^1}\phi_1(\theta )),
\eqno(6.14d)
$$
where the contour $S^1=\{\l\in \C{\bp}^1: |\l|=1\}$
circles once around $\l =0$ and  the contour integral determines
 residue $Res$ at the point $\l =0$. It follows from (6.14)
that the action of the algebra $C^1(\fu , \co^{\bf g}_\rp )$ on
$\{A_\mu\}$ is nonlocal, i.e. $\d_\theta A_{\mu}$ depend on values
of $\{A_\mu\}$ at all points $x\in U$ since $\phi_{1,2}(\theta )$
hiddenly depend on $\{A_\mu\}$.

\subsection{An analogue of the Virasoro algebra in integrable 4D CFT's}
\label{6.4}

In \S\S 5.4, 5.5, 8.2 of the paper~\cite{Po}  the local group $\fh$ of
biholomorphisms of the twistor space $\rp$ and its action on the
space of local solutions to the SDYM equations were described.
To this group there corresponds the algebra $C^0(\fu , \cv_\rp )$
of 0-cochains of the cover  $\fu =\{\U_1, \U_2\}$ of $\rp$ with
values in the sheaf $\cv_\rp$ (of germs) of holomorphic vector
fields on  $\rp =\U_1\cup\U_2$. In particular, the   algebra
$H^0(\rp , \cv_\rp )$ of global sections of the sheaf $\cv_\rp$
corresponds to biholomorphisms of $\rp$ preserving the transition
function $f_{12}$. However, these algebras are not correct
generalizations of the Virasoro algebra.

\smallskip

An analogue of the Virasoro algebra is the algebra $\cv_\rp (\U_{12})$
of holomorphic vector fields on $\U_{12}=\U_1\cap\U_2\subset\rp$.
It is a subalgebra of the algebra
$$
C^1(\fu , \cv_\rp )\simeq \cv_\rp (\U_{12})\oplus \cv_\rp (\U_{12})
\eqno(6.15)
$$
of 1-cochains of the cover $\fu$ with values in the sheaf $\cv_\rp$.
Elements of the algebra $C^1(\fu , \cv_\rp )$ are collections of
vector fields
$$
\chi =\{\chi_{12}, \chi_{21}\}=\{\chi_{12}^a\frac{\p}{\p z^a_1},
\chi_{21}^a\frac{\p}{\p z^a_2}\}
\eqno(6.16)
$$
with ordered ``cohomology indices".

\smallskip

Let us define the following action of the algebra $C^1(\fu , \cv_\rp )$
on the complex coordinates on $\rp :  \d_\chi z^a_1=\chi^a_{12},\
\d_\chi z^a_2=\chi^a_{21},$ where $\d_\chi z^a_{1,2}$ are defined
on $\U_{12}$. From the Kodaira-Spencer deformation theory~\cite{Kod}
it follows that the algebra (6.15) acts on the transition function
$f_{12}$ of the space $\rp$ by the formula
$$
\d_\chi f^a_{12}=\chi_{12}^a - \frac{\p f_{12}^a}{\p z^b_2}
\chi^b_{21}\quad \Leftrightarrow\quad
\d_\chi f_{12}:= \d_\chi f^a_{12}  \frac{\p}{\p z^a_1} =
\chi_{12}-\chi_{21}.
\eqno(6.17)
$$
The algebra  $C^0(\fu , \cv_\rp )$ acts on the transition function
$f_{12}$ of the space $\rp$  via the twisted homomorphism
$$
h^0: C^0(\fu , \cv_\rp )\ni\{\chi_1,\chi_2\}\mapsto\{\chi_1-\chi_2,
\chi_2-\chi_1\}\in C^1(\fu , \cv_\rp )
\eqno(6.18)
$$
of the algebra $C^0(\fu , \cv_\rp )$ into the algebra $C^1(\fu ,
\cv_\rp)$.

Let us consider a subalgebra
$$
C_\t^1(\fu , \cv_\rp )=\{\chi \in C^1(\fu , \cv_\rp ):
\chi_{12}=\chi_{21}\}
\eqno(6.19)
$$
of the algebra  $C^1(\fu , \cv_\rp )$. Then, the space
$$
Z^1(\fu , \cv_\rp )=\{\chi \in C^1(\fu , \cv_\rp ):
\chi_{12}=-\chi_{21}\}
$$
of 1-cocycles of the cover $\fu$ with values in $\cv_\rp$ is isomorphic
to the quotient space
$$
Z^1(\fu , \cv_\rp )\simeq  C^1(\fu , \cv_\rp )/C_\t^1(\fu , \cv_\rp ).
$$
So we can always split $2\chi\in C^1(\fu , \cv_\rp )$ in symmetric
$\chi^{sym}:=\{\chi_{12}+\chi_{21}, \chi_{12}+\chi_{21}\}\in
C_\t^1(\fu , \cv_\rp )$ and antisymmetric
$\chi^{ant}:=\{\chi_{12}-\chi_{21}, \chi_{21}-\chi_{12}\}\in
Z^1(\fu , \cv_\rp )$  parts.

\smallskip

Notice that $\d_\chi f:=\{\d_\chi f_{12}, \d_\chi f_{21}\}\in
Z^1(\fu , \cv_\rp)$, and the quotient space
$$
H^1(\fu , \cv_\rp):= h^0(C^0(\fu , \cv_\rp))\backslash Z^1(\fu , \cv_\rp)
\eqno(6.20)
$$
describes nontrivial infinitesimal deformations of the complex structure
of $\rp$. For the cover $\fu =\{\U_1,\U_2\}$ charts $\U_1,\U_2$ are Stein
manifolds, and we have $H^1(\rp , \cv_\rp)=H^1(\fu , \cv_\rp)$. In contrast
with the 2D case (6.4) we now have $H^1(\rp , \cv_\rp)\ne 0$. Hence, the
transformations (6.17) of the transition function in general change the
complex structure on $\rp$ and therefore change the conformal structure
on $U$.

\smallskip

We consider a holomorphic bundle $E'\to\rp$ which corresponds to
a self-dual bundle $E\to U$. Then $C^1(\fu,\cv_\rp )$ is the symmetry
algebra of the following {\it system} of
equations:
$$
\bar\p^2=0,\quad \bar\p B+B\wedge B=0,
\eqno(6.21)
$$
where the equations $\bar\p^2=0$ are the integrability
conditions of an almost complex structure on $\rp$, and the equations
$\bar\p_B^2=\bar\p B+B\wedge B=0$ are the conditions of holomorphy
of the bundle $E'$ over  $\rp$.

\smallskip

The symmetric part  $\chi^{sym}$ of
$\chi\in C^1(\fu,\cv_\rp )$  does not change the transition function
$f_{12}$ (see (6.17)) and the complex structure of $\rp$. Hence,
one may define the following holomorphic action of the algebra
$C^1(\fu , \cv_\rp )$ on transition matrices $\cf_{12}$ of the
holomorphic bundle $E'$:
$$
\d_\chi \cf_{12}=\chi^{sym}_{12}(\cf_{12})=
\chi_{12}(\cf_{12})+\chi_{21}(\cf_{12}).
\eqno(6.22)
$$
To define an action of this algebra on smooth
objects $B=\{B^{(1)}, B^{(2)}\}$, $\{\psi_1, \psi_2\}$ on $\rp$
and on a self-dual gauge potential $A$ on $U$, one should:
1)  substitute $\d_\chi\cf_{12}$ from (6.22) into (6.12) instead of
$\d_\theta\cf_{12}$ and obtain $\phi_1(\chi ), \phi_2(\chi )$,
2) substitute $\phi_{1,2}(\chi )$ into (6.13) instead of
$\phi_{1,2}(\theta )$, 3) substitute $\phi_{1,2}(\chi )$ into (6.14)
instead of $\phi_{1,2}(\theta )$ and obtain $\d_\chi A_{y^1}$ etc.

\subsection{Infinitesimal deformations of complex structures on $\rp$}
\label{6.5}

Recall that a conformal structure $[g]$ on $U\,\subset\,M$
is called self-dual if the Weyl tensor for any metric $g$ on $U$ in
the conformal equivalence class
$[g]$ is self-dual~\cite{AHS}. By virtue of the twistor
correspondence~\cite{Pen, AHS} the moduli space of self-dual conformal
structures on an open subset $U$ of a 4-manifold $M$ is bijective to
the moduli space of complex
structures on the twistor space $\rp$ of $U$. Thus, deformations of the
complex structure on $\rp$ are equivalent to deformations of the
conformal structure on $U$.

\smallskip

All algebras of infinitesimal symmetries of the self-dual gravity
equations known by now (see e.g.~\cite{PBR} and references therein)
are subalgebras in the algebra $C^1(\fu ,\cv_\rp)$. The action of the
algebra  $C^0(\fu ,\cv_\rp)$ transforms $f_{12}$ into an equivalent
transition function
and therefore preserves the conformal structure on $U$. At the same
time, the action of the algebra $C^0(\fu ,\cv_\rp)$ on transition
matrices of holomorphic bundles $E'\to\rp$ is not trivial.

\smallskip

To define an action of the algebra $C^1(\fu ,\cv_\rp)$ on
smooth objects on $\rp$ (functions, q-forms etc), we should
introduce:

\smallskip

1) the sheaf $\ct^{1,0}$ of (1,0) vector fields on $\rp$,
holomorphic along fibres $\cp$ of the bundle $\rp\to U$;

2) the sheaf $\cw$ of $\bar\p$-closed (0,1)-forms $W$ on $\rp$
with values in $\ct^{1,0}$,
vanishing on the distribution $V^{0,1}$ (see \S\,4.2).

\smallskip

On any open set $\U\subset\rp$ sections $W\in \cw (\U)$ of the sheaf
$\cw$ have to satisfy the equations
$$
\bar\p W =0
\eqno(6.23)
$$
Then  we have the exact sequence of sheaves
$$
0\lra\cv_\rp\lra\ct^{1,0}\lra\cw\lra 0
\eqno(6.24)
$$
and the corresponding exact sequence of cohomology spaces
$$
0\lra H^0(\rp,\cv_\rp)\lra H^0(\rp,\ct^{1,0})\lra H^0(\rp,\cw)
\lra H^1(\rp,\cv_\rp)\lra 0,
\eqno(6.25)
$$
describing infinitesimal deformations of the complex structure of the
twistor space $\rp$.

\smallskip

{}From (6.25) it follows that for any element $\d_\chi f\in
Z^1(\fu ,\cv_\rp )\subset Z^1(\fu ,\ct^{1,0})$ there exists an
element $\vp =\{\vp_1,\vp_2\}\in  C^0 (\fu ,\ct^{1,0})$ such that
$$
\d_\chi  f=\{\chi_{12}-\chi_{21}, \chi_{21}-\chi_{12}\}=\{\vp_1(\chi )
-\vp_2(\chi ), \vp_2(\chi )-\vp_1(\chi )\}\in  h^0(C^0 (\fu ,
\ct^{1,0})),
\eqno(6.26)
$$
where $h^0: \{\vp_1, \vp_2\}\mapsto \{\vp_1-\vp_2, \vp_2-\vp_1\}$
is the twisted homomorphism of the algebra $C^0(\fu ,\ct^{1,0})$
into the algebra $C^1(\fu ,\ct^{1,0})$.
Then  for infinitesimal transformations of the complex coordinates on
$\rp =\U_1\cup\U_2$ we have
$$
\tilde\d_\chi z^a_1:=\vp^a_1(\chi ;z_1,\bar z_1),\quad
\tilde\d_\chi z^a_2:=\vp^a_2(\chi ;z_2,\bar z_2).
\eqno(6.27)
$$
To preserve the reality of the conformal structure on $U$, one should
define real subalgebras of the algebras $C^1(\fu ,\cv_{\rp})$  and
$C^0 (\fu ,\ct^{1,0})$ by analogy with \S\S\,6.6,\,7.7 of~\cite{Po}.
 We shall not write down transformations of the conformal structure
on $U$ since this will require a lot of additional explanations.

{}Formula (6.26) defines a (twisted) isomorphism between the algebra
$C^1(\fu , \cv_{\rp})$   and a subalgebra in $C^0(\fu , \ct^{1,0})$.
The action (6.27) of the algebra $C^1(\fu , \cv_{\rp})$ based on this
isomorphism is not holomorphic, i.e. it changes the complex structure
of $\rp$. Having this action one can define an action of
$C^1(\fu , \cv_{\rp})$ on any object on  $\rp$. In particular, we have
the nonholomorphic action
$$
\tilde\d_\chi\cf_{12}= \tilde\d_\chi z^a_1\frac{\p\cf_{12}}{\p z^a_1}=
\vp_1(\cf_{12})= \vp_2(\cf_{12})+\chi_{12}(\cf_{12})
\eqno(6.28)
$$
of the algebra $C^1(\fu , \cv_{\rp})$ on transition matrices $\cf_{12}$
of the holomorphic bundle $E'\to\rp$. If $\vp^a_1$ and $\vp^a_2$ are
holomorphic functions on $\U_1$ and $\U_2$, respectively, then
$\{\vp_1, \vp_2\}\in C^0(\fu , \cv_{\rp})\subset C^0(\fu , \ct^{1,0})$
and formula (6.28) defines the holomorphic action of the algebra
$C^0(\fu , \cv_{\rp})$ on transition matrices $\cf_{12}$ of the
bundle $E'$ over the twistor space $\rp$.

\newpage

\section{Quantization programme}
\label{7}

\subsection{Symmetry algebras and their representations}
\label{7.1}

In \S\S\,6.3-6.5  we have described the action of the algebra
$C^1(\fu ,\cv_\rp )\dotplus C^1(\fu ,\co^{\bf g}_\rp )$
($\dotplus$ denotes the semidirect sum) on the holomorphic transition
{}function of the twistor space $\rp$ and on holomorphic transition
matrices in the bundles $E'\to\rp$. There is a subalgebra in the algebra
$C^0(\fu ,\ct^{1,0} )\dotplus C^0(\fu ,\cs^{\bf g}_\rp )$
which is isomorphic to the algebra
$C^1(\fu ,\cv_\rp )\dotplus C^1(\fu ,\co^{\bf g}_\rp )$
and is defined with the help of solutions to the infinitesimal
Riemann-Hilbert problems (6.12), (6.26).
This subalgebra acts on smooth objects defined on $\rp$. The action
of the symmetry algebra on smooth objects in four dimensions is defined
with the help of contour integrals (residue at the point $\l =0$).
We summarize the description of the symmetry algebras of CFT's in two
dimensions and their analogues in four and six dimensions in the following
table:

\medskip
\begin{center}
\begin{tabular}{|c|c|c|c|}
\hline
2D & 6D& 6D&4D\\[1mm]
 &{{\small holomorphic setting}}&
{{\small smooth setting}}&\\[1mm]
\hline
\parbox{2.5cm}{the Virasoro algebra $\Leftrightarrow
C^1(\fo ,\cv_{\C\bp^1})$}
& \parbox{2.2cm}{the algebra\\ of 1-cochains\\ $C^1(\fu ,\cv_{\rp})$}&
\parbox{2.5cm}{a subalgebra in the algebra $C^0(\fu ,\ct^{1,0})$}&
\parbox{5.3cm}{the algebra of (nonlocal)\\ transformations\\
$\delta_\chi (...)={\mathop{\mbox{Res}}\limits_{\l =0}}(...),
\chi\in C^1(\fu ,\cv_\rp)$}
\\[1cm]
\hline
\parbox{3cm}{the affine Lie alge-\\bra ${\bf g}\otimes\C[\l ,\l^{-1}]$ \\
$\Leftrightarrow C^1(\fo ,\co^{\bf g}_{\C\bp^1})$} &
 \parbox{2.2cm}{the algebra\\ of 1-cochains\\ $C^1(\fu ,\co^{\bf g}_{\rp})$}&
\parbox{2.5cm}{a subalgebra in the algebra $C^0(\fu ,\cs^{\bf g}_\rp)$}&
\parbox{5.3cm}{the algebra of (nonlocal)\\ transformations\\
$\delta_\theta (...)={\mathop{\mbox{Res}}\limits_{\l =0}}(...),
\theta\in C^1(\fu ,\co^{\bf g}_\rp)$}
\\[1cm]
\hline
\end{tabular}
\end{center}
\smallskip

Notice that the subalgebra $C^0(\fu ,\cv_\rp )$ of the algebra
$C^1(\fu ,\cv_\rp )$ also plays an important role. The action of this
algebra does not change the complex structure on $\rp$ and therefore
does not change the conformal structure  on $U\subset M$. In many
particular cases of a choice of a 4-manifold $M$, the subalgebra
$H^0(\rp ,\cv_\rp )$ of this algebra is finite-dimensional and
coincides with the algebra of the conformal group in the ``flat limit".
As a first step, fields of any 4D CFT should be classified using
representations of this algebra.

\smallskip

Recall that at a quantum level, rational 2D CFT's are solvable for
energy spectrum and correlation functions. Moreover, in genus zero
the quantization of these theories in fact amounts to the construction of
representations of the Virasoro and affine Lie algebras, and the Ward
identities uniquely determine the correlation functions for descendants
of the identity operator.

\smallskip

Analogous results can be expected for integrable 4D CFT's. Indeed,
{}for the basic representative of these models - the SDYM model on an
open ball $U\subset\R^4$ - the moduli space is the double coset space
(4.13) and the algebra $C^1(\fu , \co^{\bf g}_\rp )$ is the
complexification of the Lie algebra
of the symmetry group $C^1_\tau (\fu ,\ch )$ from (4.13). Hence,
as in the case of the chiral 2D WZNW theory on a disk,
symmetries completely determine the phase space of the SDYM model on
$U$ and quantization of this model amounts to  constructing
representations of the algebra $C^1(\fu ,\co^{\bf g}_\rp )$ and to
choosing in
them the subspaces which are invariant w.r.t. the action of the
subalgebra $C^0(\fu ,\co^{\bf g}_\rp )$ of $C^1(\fu ,\co^{\bf g}_\rp )$.
Extensions of all
these algebras arise only as quantum effects of normal ordering.

\subsection{Quantization of the SDYM model}
\label{7.2}

In this section we want to discuss quantization of integrable
4D CFT's.
Remembering the connection between 2D
CFT's and  ordinary 3D CS theories, one may come to the reasonable
conclusion that the quantization of integrable 4D CFT's may be much
more successful if we use their connection with holomorphic CSW theories.

\smallskip

When quantizing the holomorphic CSW theory on twistor spaces,
one may use the results on the quantization of the ordinary CS theory
(see e.g.~\cite{Wit1}-\cite{EMSS},\cite{DJT}  and references therein)
after a proper generalization. We are mainly interested in (canonical)
quantization of the SDYM model on $U$ since this model is the
closest anologue of the chiral 2D WZNW model on a disk.
As such, we have to put $\hat B_3=0$ in eqs.(3.1) on the twistor space
$\rp$ of $U$, which leads
to the equations (cf.(4.6))
$$
\bar\p B + B\wedge B=0
\eqno(7.1)
$$
equivalent to the SDYM equations on $U$, as has been discussed in this
paper. The comparison with the ordinary CS theory in the Hamiltonian
approach shows that the coordinate $\bar\l$ conjugate to $\l\in\C\bp^1$
may be considered as (complex) time of the holomorphic CSW theory.

\smallskip

Some problems related to quantization of the SDYM model were
discussed in~\cite{NS, LMNS, CY}. The quantization was carried out
in four dimensions in terms of $\fg$-valued  fields $A_\mu$ or in terms
of a $G$-valued scalar field by using the Yang gauge. But the obtained
results are fragmentary; the picture is not complete and far from what
we have in 2D CFT's.

\smallskip

Here we shall discuss the quantization of the SDYM model using
its connection with the holomorphic CSW theory.
In quantization of constrained systems one can use two standard
approaches:
1) one first solves the constraints and then performs the quantization
of the moduli space;
2) one first quantizes the free theory and then imposes (quantum)
constraints. Both approaches will be discussed. We shall
write down the list of questions and open problems whose solutions are
necessary to give the holomorphic CSW and the SDYM theories a status
of quantum field theories.

\smallskip

1. One should rewrite a symplectic structure $\hat\o$ on the space
of gauge potentials or their relatives in
terms of fields on the twistor space $\rp$. This 2-form $\hat\o$
induces a symplectic structure $\tilde\o$ on the moduli space $\cm_U$ of
solutions to eqs.(7.1), and the cohomology class $[\tilde\o ]\in
H^2(\cm_U,\R)$ has to be integral.

\smallskip

2. Over the moduli space $\cm_U$ one should define a complex
line bundle $\cl$ with the Chern class $c_1(\cl)=[\tilde\o]$.
Then $\cl$ admits a connection with the curvature  2-form equal
to $\tilde\o$.

\smallskip

3. A choice of a conformal  structure $[g]$ on $U$ induces the
complex structure $\cj$ on the twistor space $\rp$ and
endows the moduli space $\cm_U$ with a complex structure
which we shall denote by the same letter $\cj$. Then the bundle $\cl$ over
$(\cm_U,\cj)$ has a holomorphic structure, and a quantum
Hilbert space
of the SDYM theory can be introduced as the space $H_\cj$ of (global)
holomorphic sections of $\cl$.

\smallskip

4. Is it possible to introduce the bundle $\cl\to\cm_U$ as the
holomorphic determinant line bundle Det$\,\bar\p_B$ of the operator
$\bar\p_B=\bar\p + B$ on $\rp$?

\smallskip

5. The action functional of the holomorphic CSW theory on a Calabi-Yau
3-fold has a simple form (3.2) analogous to the action of the
standard CS theory. How should one modify this action if we go over
to the case of an arbitrary complex 3-manifold?

\smallskip

6. One should lift the action of the symmetry groups and algebras
described in this paper up to an action on the space $H_\cj$ of
holomorphic sections of the bundle $\cl$ over $\cm_U$. What is an
extension (central or not) of these groups and algebras? Finding
an extension $\hat C^1(\fu,\co^{\bf g}_\rp)$ of the algebra
$C^1(\fu,\co^{\bf g}_\rp)$ is equivalent
to finding a curvature of the bundle $\cl$ since this curvature
represents a local anomaly. It is also necessary to find an extension
$\hat C^1(\fu, \cv_\rp)$ of the algebra $C^1(\fu, \cv_\rp)$.

\smallskip

7. What can be said about representations of the algebras
$C^1(\fu,\cv_\rp)$ and $C^1(\fu,\co^{\bf g}_\rp)$? Which of these
representations are connected with the Hilbert space $H_\cj$?

\smallskip

8. In the quantum hCSW and SDYM theories there exist
Sugawara-type formulae, i.e. generators of the algebra
$\hat C^1(\fu,\cv_\rp)$ can be expressed in terms of
generators of the algebra $\hat C^1(\fu,\co^{\bf g}_\rp)$.
This follows from the fact that any transformation of transition
matrices of a holomorphic bundle $E'\to\rp$ under the
holomorphic action of the algebra
$C^1(\fu,\cv_\rp)$ can be compensated by an action of the algebra
$C^1(\fu,\co^{\bf g}_\rp)$. What are the explicit formulae
connecting the generators of these algebras?

\smallskip

9. One should write down Ward identities resulting from the symmetry
algebra  $C^1(\fu,\cv_\rp)\dotplus C^1(\fu,\co^{\bf g}_\rp)$. To what
extent do these identities define correlation functions?

\smallskip

\noindent
Clearly, to carry out this quantization programme, it will be
necessary to overcome a number of technical difficulties.

\smallskip

{\bf Remarks}

\smallskip

1. It is known that on the moduli space $\cm_U$ of solutions to the
SDYM equations on $U$ one can introduce the hyperK\"ahler structure
\cite{It}. On the other hand, in~\cite{Po} it has been shown that $\cm_U$
is the double coset space (4.13). There is no contradiction between these
structures. In fact, Joyce~\cite{Jo} has shown that on many double coset
spaces one can introduce 3 complex structures.

\smallskip

2. In the finite-dimensional case, any double coset space has the form
$K\bl G/H$, where groups $G, H$ and $K$ are finite-dimensional.
Quantization of a homogeneous space $G/H$ is well known (orbit
method~\cite{Ki}, geometric quantization~\cite{Woo}) and is reduced to
the construction of representations of the group $G$ on polarized sections
of a complex line bundle over $G/H$. In the chiral 2D WZNW model
this approach is used for quantization of the homogeneous space $LG/G$
of the loop group $LG$. By quantizing the biquotient space $K\bl G/H$,
we first have to associate a representation space of the group $G$
with the homogeneous space $G/H$ (this is standard) and then
choose in it subspaces which are invariant under the action of the
group $K$. Quantization of the biquotient moduli space $\cm_U$
will be an infinite-dimensional variant of this construction.

\subsection{The analytic geometry of integrable 4D CFT's}
\label{7.3}

Let us introduce an almost complex structure on a real 6-manifold
$\cz$, that is equivalent to assigning the operator $\bar\p$ on
$\cz$. Integrability conditions of this almost complex structure
are equivalent to the equations
$$
\bar\p^2=0
\eqno(7.2a)
$$
and different complex structures on $\cz$ correspond to different
operators $\bar\p$.

\smallskip

Having the complex 3-manifold $(\cz ,\bar\p )$, let us consider a
complex vector bundle $E'$ over $\cz$. Assignment of an almost
complex structure on $E'$ is equivalent to the introduction of
the (0,1)-component $\bar\p_{\hat B}=\bar\p +\hat B$ of connection on
$E'$. Integrability of this almost complex structure is
equivalent to the validity  of equations
$$
\bar\p_{\hat B}^2=0\ \Leftrightarrow \
\bar\p\hat B +\hat B\wedge\hat B=0.
$$
In particular, we shall be interested in the case when $\cz$  is
the twistor space of a self-dual 4-manifold $M$ and (0,1)-connections
$\bar\p_{B}=\bar\p + B$ are defined on a holomorphic bundle
$E'\to\cz$, which is connected with a self-dual bundle $E\to M$. For
such connections we have the equations
$$
\bar\p_{B}^2=0\ \Leftrightarrow \
\bar\p B + B\wedge B=0,
\eqno(7.2b)
$$
which are equivalent, as was discussed, to the SDYM equations on $M$.

\smallskip

The general picture arising as a result of quantization of the SDYM
model on a self-dual 4-manifold $M$ and of the holomorphic CSW theory on
the twistor space $\cz$ of $M$ resembles the one that arises in the
quantization of 2D CFT's and ordinary CS theories and is as follows.
Let $[g]$ be a self-dual conformal structure on a 4-manifold $M$ and
let $\cj$ be a complex structure on the twistor space $\cz$ of $M$.
As has already been noted, there exists a bijection~\cite{Pen, AHS}
between the moduli space of self-dual conformal structures on $M$ and
the moduli space $\fx$ of complex structures on $\cz$. Note that
energy-momentum tensors of integrable 4D CFT's on $M$ with a metric
$g$ are connected with infinitesimal variations of the metric $g$ on $M$
and, as we have said above,  variations of the
metric on $M$ are proportional to variations of the complex structure
$\cj$ on the twistor space $\cz$ of $M$. Notice that the
energy-momentum tensor for the SDYM model can be obtained by using the
Donaldson-Nair-Schiff action~\cite{Do, NS}. It seems that this
has not been done in the literature.

\smallskip

A choice of a complex structure $\cj$ on the twistor space $\cz$
endows the moduli space of solutions to the SDYM equations on $M$ with
a complex structure which we shall denote by the same letter $\cj$.
We denote this moduli space with the complex structure $\cj$
by $\cm_\cj$. Then one can introduce a bundle
$$
\tilde \cm\to\fx
\eqno(7.3)
$$
with fibres $\cm_\cj$ at the points $\cj\in\fx$. The total space
$\tilde\cm$ of
this bundle is the moduli space of solutions to eqs.(7.2). In other
words, this is the moduli space of pairs $(\cz , E')$, where $\cz$ is
a complex 3-manifold and $E'$ is a holomorphic bundle over $\cz$.
Study of this moduli space is important for understanding
extended mirror symmetries.

\smallskip

{\bf Remark.} If we consider holomorphic bundles $E'$ over Calabi-Yau
3-folds $\cz$, the moduli space $\fx$ of complex structures on
$\cz$ should be replaced by a ``larger" moduli space $\hat\fx$ of
pairs (a complex structure $\cj$ on $\cz$, a K\"ahler structure
$\o$ on $\cz$). The moduli space $\cm_{\cj ,\o}$ of
{\it stable} holomorphic bundles depends not only on the complex
structure $\cj$ on $\cz$, but also on the K\"ahler structure
$\o$ on $\cz$. Then, instead of the bundle (7.3) one should
consider a bundle
$$
\hat \cm\to\hat\fx
\eqno(7.4)
$$
with fibres $\cm_{\cj ,\o}$ at the points $(\cj ,\o )\in
\hat\fx $. One should use the total space $\hat \cm$ of the bundle (7.4)
in the consideration of the extended mirror symmetry between CY 3-folds
with holomorphic bundles.

\smallskip

{}Further, let $\cm_\cj$ be the moduli space of
solutions to the SDYM equations on $M$ and let $H_\cj$ be the quantum
Hilbert space of holomorphic sections of the line bundle $\cl$ over
$\cm_\cj$. The space $H_\cj$ depends on $\cj\in \fx$ and one can
introduce a holomorphic vector bundle
$$
p: \tilde H\lra \fx
\eqno(7.5)
$$
with fibres $H_\cj$ at the points $\cj\in\fx$. Then one may raise the
question about the existence of a (projectively) flat connection on
the bundle (7.5).  If such a connection exists, then as the quantum
Hilbert space one may take the space of covariantly constant sections
of the vector bundle $\tilde H$.  Thus, the Friedan-Shenker
approach~\cite{FS} to 2D CFT's  can be generalized to integrable
4D CFT's. In this geometric approach the energy-momentum tensor of an
integrable  4D CFT will induce a (flat) connection on the bundle
(7.5), and equations defining covariantly constant sections of the
bundle (7.5) will be equivalent to Ward identities.  Moreover,
it can be expected that for CFT's on an open set $U\subset M$,
an (n+1)-point amplitude with energy-momentum insertion will be related
to an n-point amplitude and its derivative w.r.t. moduli parameters from
the space $\fx$ because of the Ward identities associated to the
algebra $\hat C(\fu,\cv_\rp )$ (cf.~\cite{egu}).
Analogously, the Ward identities for the algebra  $\hat
C(\fu,\co^{\bf g}_\rp )$ will involve derivatives w.r.t. moduli
parameters from $\cm_\cj$. The described symmetries can essentially
simplify the study of integrable 4D conformal quantum field theories.

\newpage

\section{Conclusion}
\label{8}

\smallskip

In this paper we studied the moduli spaces of solutions to the field
equations of the ordinary  Chern-Simons theory, holomorphic
Chern-Simons-Witten theory and  the SDYM theory. Integrable 4D CFT's,
their connection with holomorphic CSW theories, parallels with 2D
CFT's and symmetries have been described. We have also discussed the
programme of quantization of the SDYM model on a self-dual 4-manifold
$M$ based on the equivalence of this model to a subsector of the
holomorphic CSW model on the twistor space $\cz$ of $M$.

\smallskip

One of the
purposes of this paper was the advance of integrable 4D CFT's as
perspective candidates for the role of 4D quantum field theories with
vanishing beta functions. It was argued that integrable conformal field
theories in four dimensions can be developed to such an extent
of generality as free and rational conformal field theories in two
dimensions. The main argument is the invariance of integrable 4D
CFT's under the action of the infinite-dimensional algebras, which will put
severe restrictions on the Green functions through the Ward  identities
and will strongly constraint underlying field theories. Much work remains
to be done.

\smallskip

\section*{Acknowledgements}
\addtocontents{toc}{\medskip}
\addcontentsline{toc}{appe}{\noindent\bf Acknowledgements\hfill
\medskip
}

The author is grateful to Yu.I.Manin and I.T.Todorov for helpful
discussions. I also thank the Max-Planck-Institut f\"ur Mathematik in
Bonn, where part of this work was done, for its hospitality and the
Alexander von Humboldt Foundation for support.  This work is supported
in part by the grant RFBR-98-01-00173.

\section*{Appendix A. Sheaves of groups}
\addcontentsline{toc}{appe}{\\
\medskip
\bf Appendix A. Sheaves of  groups\hfill}

Let us consider a topological space $X$ and recall the definitions of a
presheaf and a sheaf of groups over $X$ (see e.g.~\cite{GR, Hir}).

\smallskip

One has  a {\it presheaf} $\{\fs (U), r^U_V\}$ of groups over a
topological space $X$ if with any nonempty
open set $U$ of the space $X$ one associates a group $\fs(U)$
and  with any two open sets $U$ and $V$ with $V\subset U$
one associates a homomorphism $r^U_V: \fs(U)\to\fs(V)$ satisfying
the following conditions:
(i) the homomorphism  $r^U_U: \fs(U)\to\fs(U)$ is the identity
map id$_U$;
(ii) if $W\subset V\subset U$, then $r^U_W=r^V_W\circ r^U_V$.

\smallskip

A {\it sheaf} of groups over a topological space $X$ is a topological
space $\fs$ with a {\it local homeomorphism} $\pi : \fs\to X$. This
means that any point $s\in\fs$ has an open neighbourhood $V$ in $\fs$
such that $\pi (V)$ is open in $X$ and $\pi : V\to\pi (V)$ is a
homeomorphism.  A set $\fs_x=\pi^{-1}(x)$ is called a {\it stalk} of
the sheaf $\fs$ over $x\in X$, and the map $\pi$ is called the
projection.  For any point $x\in X$ the stalk $\fs_x$ is a group, and
the group operations are continuous.

\smallskip

A {\it section}  of a sheaf $\fs$ over an open set $U$ of the space $X$
is a continuous map $s: U\to\fs$ such that $\pi\circ s=$id$_U$. A set
$\fs (U):=\Gamma (U,\fs)$ of all sections of the sheaf $\fs$ of groups
over $U$ is a group. Corresponding to any open set $U$ of the space
$X$ the group $\fs(U)$ of sections of the sheaf $\fs$ over $U$ and to
any two open sets $U,V$ with $V\subset U$ the restriction homomorphism
$r^U_V: \fs(U)\to\fs(V)$, we obtain the presheaf $\{\fs(U),r^U_V\}$ over
$X$. This presheaf is called the canonical presheaf.

\smallskip

On the other hand, one can associate a sheaf with any presheaf
$\{\fs(U),r^U_V\}.$
Let
$$
\fs_x={\mathop{\mbox{lim}}\limits_{\stackrel{\lra}{x\in U}}}\fs(U)
$$
be a direct limit of sets $\fs(U)$. There exists a natural map
$r^U_x : \fs(U)\to\fs_x$, $x\in U$, sending elements from
$\fs(U)$ into their equivalence classes in the direct limit.
If $s\in\fs(U)$, then
${\bf s}_x:=r^U_x(s)$ is called a {\it germ} of the section $s$ at the
point $x$, and $s$ is called a {\it representative} of the germ ${\bf s}_x$.
Put another way, two sections $s,s'\in \fs(U)$ are called {\it equivalent}
at the point $x\in U$ if there exists an open neighbourhood $V\subset U$
such that $s|_V=s'|_V$; the equivalence class of such sections is called
the germ ${\bf s}_x$ of section $s$ at the point $x$. Put
$$
\fs={\mathop\cup\limits_{x\in X}}\fs_x
$$
and let $\pi : \fs\to X$ be a projection mapping points from $\fs_x$
into $x$. The set $\fs$ is
equipped with a topology, the basis of open sets of which consists of sets
$\{{\bf s}_x, x\in U\}$ for all possible $s\in\fs(U), U\subset X$.
In this topology $\pi$ is a local homeomorphism, and we
obtain the sheaf $\fs$.

\smallskip

Let $X$ be a smooth manifold. Consider a complex (non-Abelian) Lie group
${\bf G}=G^\C$ and define a presheaf $\{\hat\cs(U),r^U_V\}$ of groups by
putting
$$
 \hat\cs(U):=\{C^\infty\mbox{-maps}\ f: U\to {\bf G}\},
\eqno(A.1)
$$
and using the canonical restriction homomorphisms $r^U_V$ when for
$f\in\hat\cs(U)$ its image $r^U_V(f)$ equals $f|_V\in\hat\cs(V)$,
$V\subset U$. To each elements $\a_x$ and $\b_x$ from
$\hat\cs_x:=r^U_x(\hat\cs(U))$ one can put into correspondence
their pointwise
multiplication $\a_x\b_x$.  To this presheaf $\{\hat\cs(U), r^U_V\}$
there corresponds the sheaf $\hat\cs$ of germs of smooth maps of the
space $X$ into the group $\bf G$.

\smallskip

Suppose now that $X$ is a complex manifold. Then one can define a
presheaf $\{\hat\ch(U), r^U_V\}$ of groups assuming that
$$
\hat\ch(U)\equiv\co^{\bf G}(U):=\{\mbox{holomorphic maps}\
h: U\to {\bf G}\},
\eqno(A.2)
$$
and associate with it the sheaf $\hat\ch\equiv\co^{\bf G}$ of germs
of holomorphic maps of the space $X$ into the complex Lie group $\bf G$.

\section*{Appendix B. Cohomology sets and vector bundles}
\addcontentsline{toc}{appe}{\\
\medskip
\bf Appendix B. Cohomology sets and vector bundles\hfill}

We consider a complex manifold $X$ and a sheaf $\fs$ coinciding
with either the sheaf $\hat\cs$ or the sheaf $\hat\ch$ introduced in
Appendix A. One can also consider a sheaf $\fs$ of Abelian groups
with addition as a group operation.

\smallskip

\v{C}ech cohomology sets $H^0(X,\fs)$ and $H^1(X,\fs)$ of the space $X$
with values in the sheaf $\fs$ of groups are defined as follows~\cite{GR,
Hir}.

\smallskip

Let there be given an open cover $\fu=\{\U_\a\}$, $\a\in I$, of the
manifold $X$.
The family $\la\U_0,...,\U_q\ra$ of elements of the cover such that
$\U_0\cap...\cap\U_q\ne \varnothing$ is called a q-simplex. The support
of this simplex
is $\U_0\cap...\cap\U_q$. Define a {\it 0-cochain with coefficients
in} $\fs$ as a map $f$: $\la\a\ra\mapsto f_\a$, where $\a\in I$ and
$f_\a$ is a section of the sheaf $\fs$ over $\U_\a$,
$$
f_\a\in\fs(\U_\a):=\Gamma (\U_\a, \fs).
\eqno(B.1)
$$
A set of 0-cochains is denoted by $C^0(\fu,\fs)$ and is a group
under the pointwise multiplication (or addition in the Abelian case).

\smallskip

Consider now the {\it ordered set} of two indices $\la\a,\b\ra$
such that $\a,\b\in I$
and $\U_\a\cap\U_\b\ne\varnothing$. Define a {\it 1-cochain} with
the coefficients in $\fs$ as a map $f$: $\la\a ,\b\ra\mapsto f_{\a\b }$,
where  $f_{\a\b }$ is a section of the sheaf $\fs$ over $\U_\a\cap\U_\b$,
$$
f_{\a\b}\in\fs(\U_\a\cap\U_\b):=\Gamma (\U_\a\cap\U_\b,\fs).
\eqno(B.2)
$$
A set of 1-cochains is denoted by $C^1(\fu,\fs)$ and is a group under the
pointwise multiplication (or addition in the Abelian case).

\smallskip

Subsets of {\it cocycles} $Z^q(\fu,\fs)\subset C^q(\fu,\fs)$ for $q=0,1$
are defined by the formulae
$$
Z^0(\fu,\fs)=\{f\in C^0(\fu,\fs): f_\a f_\b^{-1}=1\ \mbox{on}\
\U_\a\cap\U_\b\ne\varnothing\},
\eqno(B.3)
$$
$$
Z^1(\fu,\fs)=\{f\in C^1(\fu,\fs): f_{\b\a}=f_{\a\b}^{-1}\ \mbox{on}\
\U_\a\cap\U_\b\ne \varnothing ;\
f_{\a\b}f_{\b\g}f_{\g\a}=1\ \mbox{on}\
\U_\a\cap\U_\b\cap\U_\g\ne\varnothing\}.
\eqno(B.4)
$$
Notice that if one considers a sheaf $\fs$ of Abelian groups,
the cocycle conditions (B.4) should be replaced by the conditions
$f_{\a\b}+f_{\b\a}=0$,
$f_{\a\b}+f_{\b\g}+f_{\g\a}=0$ and analogously in all other definitions.
It follows from (B.3) that $Z^0(\fu,\fs)$  coincides with the group
$H^0(X,\fs):=\fs(X)\equiv\Gamma (X,\fs )$ of global sections of the
sheaf $\fs$. The set $Z^1(\fu,\fs)$ is not in general a subgroup
of the group $C^1(\fu,\fs)$. It contains the marked element $\bf 1$,
represented by the 1-cocycle $f_{\a\b}= 1$ for any $\a,\b$ such that
$\U_\a\cap\U_\b\ne\varnothing$.

\smallskip

{}For $h\in C^0(\fu,\fs)$, $f\in Z^1(\fu,\fs)$ let us define an action
$\r_0$ of the group $C^0(\fu,\fs)$ on the set $Z^1(\fu,\fs)$ by
the formula
$$
\r_0(h,f)_{\a\b}=h_\a f_{\a\b}h_{\b}^{-1}.
\eqno(B.5)
$$
So we have a map $\r_0: C^0\times Z^1\ni (h,f)\mapsto\r_0(h,f)\in Z^1$.
A set of orbits of the group $C^0$ in $Z^1$ is called a {\it 1-cohomology
set} and denoted by $H^1(\fu,\fs)$. In other words, two cocycles
$f,\tilde f\in Z^1$ are called equivalent, $f\sim\tilde f$, if
$$
\tilde f= \r_0(h,f)
\eqno(B.6)
$$
{}for some $h\in C^0$, and the 1-cohomology set $H^1=\r_0(C^0)\bl Z^1$
one calls a set  of equivalence classes of 1-cocycles. Finally, we
should take the direct limit of these sets $H^1(\fu,\fs)$ over
successive refinement of the cover $\fu$ of $X$ to obtain
$H^1(X,\fs)$, the 1-cohomology set of $X$ with the coefficients in $\fs$.
In fact, one can always choose a cover $\fu=\{\U_\a\}$ such that it
will be $H^1(\fu,\fs)=H^1(X,\fs)$ and therefore it will not be
necessary to take the direct limit of sets.  This is realized, for
instance, when the coordinate charts $\U_\a$ are Stein manifolds (see
e.g.~\cite{GR}).

\smallskip

Now consider the case when $\fs$ is the sheaf of germs of
(smooth or holomorphic) functions with values in the complex Lie
group $\bf G$. Suppose we are given a representation of $\bf G$ in
$\C^n$. It is well known that any
1-cocycle $\{f_{\a\b}\}$ from $Z^1(\fu,\fs)$ defines a unique complex
vector bundle $E'$ over $X$, obtained from the direct products
$\U_{\a}\times\C^n$  by glueing with the help of $f_{\a\b}\in {\bf
G}$. Moreover, two 1-cocycles define isomorphic complex vector
bundles over $X$ if and only if the same element from $H^1(X,\fs)$
corresponds to them. Thus, we have a one-to-one correspondence
between the set $H^1(X,\fs)$ and the set of equivalence classes of
complex vector bundles of the rank $n$ over $X$. Smooth bundles are
parametrized by the set $H^1(X,\hat\cs)$ and holomorphic bundles are
parametrized by the set $H^1(X,\hat\ch)$, where the sheaves $\hat\cs$ and
$\hat\ch$ were described in Appendix A. For more details see
e.g.~\cite{GR, Hir}.


\end{document}